\documentclass[prc,twocolumn,nofootinbib]{revtex4}
\usepackage{amsmath,amssymb,epsfig,bm,mathrsfs}
\usepackage{feynmp,slashed,color,mathrsfs,nicefrac,exscale}
\usepackage[colorlinks=true,linkcolor=blue,citecolor=blue,urlcolor=blue]{hyperref}
\newcommand{\bea}{\begin{eqnarray}}
\newcommand{\eea}{\end{eqnarray}}
\newcommand{\be}{\begin{equation}}
\newcommand{\ee}{\end{equation}}

\newcommand{\vecp}{{\bm p}}

\newcommand{\Tr}{{\rm Tr}}

\newcommand{\ie}{{i.e.}}

\def\XXint#1#2#3{{\setbox0=\hbox{$#1{#2#3}{\int}$}
     \vcenter{\hbox{$#2#3$}}\kern-.5\wd0}}

\def\prc{Phys. Rev. C}%
\def\prd{Phys. Rev. D}%
\def\physrep{Phys. Rep.}%
\definecolor{red}{rgb}{0.8,0,0}
\definecolor{violet}{rgb}{0.4,0,0.4}
\definecolor{green}{rgb}{0,0.5,0.0}
\definecolor{navy}{rgb}{0.0,0.0,0.6}
\definecolor{orange}{rgb}{0.8,0.2,0.0}
   
\begin{document}
\title{
Transport coefficients of  two-flavor quark matter from the Kubo formalism
}
\author{Arus Harutyunyan}\thanks{arus@th.physik.uni-frankfurt.de}
\affiliation{Institute for Theoretical 
   Physics, Goethe University, D-60438 Frankfurt am Main, Germany 
}

\author{Dirk H.\ Rischke}\thanks{drischke@th.physik.uni-frankfurt.de}
\affiliation{Institute for Theoretical 
   Physics, Goethe University, D-60438 Frankfurt am Main, Germany 
}

\author{Armen Sedrakian} \thanks{sedrakian@fias.uni-frankfurt.de} 
\affiliation{ Frankfurt Institute for Advanced Studies, D-60438
  Frankfurt am Main, Germany } 
\affiliation{Institute for Theoretical
  Physics, Goethe University, D-60438 Frankfurt am Main, Germany }

\begin{abstract}
  The transport coefficients of quark matter at nonzero chemical potential and temperature are
  computed within the two-flavor Nambu--Jona-Lasinio model. We apply the Kubo formalism to obtain the
  thermal ($\kappa$) and electrical ($\sigma$) conductivities as well
  as an update of the shear viscosity ($\eta$) by evaluating the
  corresponding equilibrium two-point correlation functions to 
  leading order in the $1/N_c$ expansion.  The Dirac structure of the
  self-energies and spectral functions is taken into account as these
  are evaluated from the meson-exchange Fock diagrams for
  on-mass-shell quarks.  We find that the thermal and electrical
  conductivities are decreasing functions of temperature and density
  above the Mott temperature $T_{\rm M}$ of dissolution of mesons into
  quarks, the main contributions being generated by the temporal and
  vector components of the spectral functions. The coefficients show a
  universal dependence on the ratio $T/T_{\rm M}$ for different
  densities, \ie, the results differ by a chemical-potential dependent
  constant. We also show that the Wiedemann-Franz law for the ratio
  $\sigma/\kappa$ does not hold.  The ratio $\eta/s $, where $s$ is
  the entropy density, is of order of unity (or larger) close to the Mott
  temperature and, as the temperature
  increases, approaches the AdS/CFT bound $1/4\pi$. It is also conjectured that the ratio $\kappa T/c_V $,
  with $c_V$ being the specific heat, is bounded from below by 
  $1/18$.
\end{abstract}

\maketitle

\section{Introduction}
\label{sec:introduction}

The transport coefficients of strongly interacting matter in
the regime where quarks are liberated to form an interacting
quark-gluon plasma are of interest in a number of contexts. The
high-temperature and low-density regime of the phase diagram of
deconfined QCD matter is explored by heavy-ion collision
experiments at RHIC and LHC, the collective dynamics of which is well described by
hydrodynamical models with an extremely low value of the shear
viscosity~\cite{2005NuPhA.750...30G,2006PhRvC..73f4903B,2008PhRvC..77f4901S,2008PhRvC..78c4915L,2009PhRvL.103z2302L,2011PhRvL.106s2301S,2013JPhCS.455a2044H,2006PhRvL..97o2303C,2011PhRvL.106u2302N}. The
high-density and low-temperature regime of the phase diagram is of
great interest in the astrophysics of compact stars, where transport
coefficients of deconfined QCD matter are an important input in modelling an array of astrophysical
phenomena~\cite{2002PhRvC..66a5802S,2005JHEP...09..076M,2014PhRvC..90e5205A,2017arXiv170100010S}.
The intermediate regime of moderately dense and cold deconfined QCD
matter, which is targeted by the FAIR program at
GSI~\cite{friman2011cbm} and the NICA facility at JINR~\cite{refId0},
provides a further motivation for studies of the transport coefficients in
moderately dense QCD matter close to the chiral phase-transition line.

The nonperturbative nature of QCD in the phenomenologically
interesting regimes mentioned above precludes the computation of the
transport coefficients in full QCD, therefore effective models that
capture its low-energy dynamics are required.  In this work we use the
Nambu--Jona-Lasinio (NJL)
model~\cite{1961PhRv..122..345N,1961PhRv..124..246N}, which provides a
well-tested framework of low-energy QCD for studies of vacuum and
in-matter properties of ensembles of
quarks~\cite{1991PrPNP..27..195V,1992RvMP...64..649K,2005PhR...407..205B}. Because
it captures the dynamical chiral symmetry-breaking feature of QCD it
is most suited for the studies of transport coefficients in the vicinity of the
chiral phase transition, where the elementary processes contributing
to the scattering-matrix elements are dominated by mesonic
fluctuations.

The transport formalism based on the Boltzmann equation for the quark
distribution functions can be applied to strongly interacting
ensembles in the limit where the quasiparticle concept is applicable;
in that case the collision integral is dominated by two-body
collisions between quarks moving in a mean-field between collisions.
In this work we use the Kubo-Zubarev
formalism~\cite{1957JPSJ...12..570K,zubarev1997statistical} as an alternative, in which
the transport coefficients are computed from equilibrium correlation functions at nonzero
temperature and density. It provides a 
general framework valid for a strongly interacting system, which is
characterized by nontrivial spectral functions, but requires a
resummation of an infinite series of diagrams in order to obtain the correct
scaling of the transport coefficients with the coupling (even in the
weak-interaction regime).

The understanding of the elliptic flow observed at heavy-ion collider
experiments in terms of dissipative hydrodynamics, in particular the
description of the elliptic flow by a low shear viscosity-to-entropy density ratio of the deconfined
quark phase stimulated extensive studies of the shear viscosity of
strongly interacting matter. Transport coefficients of QCD matter have
been investigated using various methods including perturbative
QCD~\cite{1994PhRvD..49.4739H,2000JHEP...11..001A,2003JHEP...05..051A,2008PhRvD..78g1501H,2013PhRvD..88h5039C,2014PhRvD..90i4014G,2016arXiv161006839H,2016arXiv161006818H},
equilibrium correlation functions within the Kubo
formalism~\cite{2005hep.ph....1284D,2008JPhG...35c5003I,2008EPJA...38...97A,LW14,LKW15,LangDiss,2016PhRvC..93d5205G,2014PhLB..734..157Q},
transport simulations of the Boltzmann
equation~\cite{2008PhRvL.100q2301X,2013PhRvC..88d5204M,2014PhRvD..90k4009P,2012PhRvC..86e4902P},
relaxation-time approximation to the Boltzmann
equation~\cite{1985PhRvD..31...53D,2010NuPhA.845..106K,2012EPJA...48..142H,2013PhRvC..88f8201G},  lattice
methods~\cite{2007PhRvD..76j1701M,2009NuPhA.830..641M,
2007PhRvL..99b2002A,2013PhRvL.111q2001A,2015JHEP...02..186A,2017arXiv170102266A}
and holographic methods~\cite{2005PhRvL..94k1601K,2008PhRvL.101m1601G,2015PhRvL.115t2301R,2014PhRvD..89j6008F,2015JHEP...06..046L,2015arXiv151003321I}  with the main
emphasis on the low values of the shear viscosity-to-entropy density ratio indicated by the
hydrodynamical modeling of heavy-ion collision
experiments~\cite{2005NuPhA.750...30G,2006PhRvC..73f4903B,2008PhRvC..77f4901S,2008PhRvC..78c4915L,2009PhRvL.103z2302L,2011PhRvL.106s2301S,2013JPhCS.455a2044H,2006PhRvL..97o2303C,2011PhRvL.106u2302N}.
The shear viscosity of quark matter was computed within the NJL model
using the Kubo formalism by several authors~\cite{2008JPhG...35c5003I,
  2008EPJA...38...97A,LW14,LKW15,LangDiss,2016PhRvC..93d5205G}.  The problem of
the resummation of an infinite series of loops required to obtain the
finite-temperature correlation functions of quark matter is simplified
by applying a $1/N_c$ power-counting
scheme~\cite{1994PhRvC..49.3283Q}, where $N_c$ is the number of colors.
At leading order the resummation then reduces to
keeping a single-loop diagram with full (dressed) propagators.  
Close to the chiral phase transition  the quark self-energies
are dominated by processes involving mesonic fluctuations, which
can be obtained within the NJL model consistent with the two-point
correlation functions.

In this work we compute the transport coefficients of quark matter at nonzero temperature
and density within the NJL model and the Kubo-Zubarev
formalism. Our main results concern the electrical and thermal
conductivities of quark matter within the setup appropriate for
relativistic quantum
fields~\cite{1984AnPhy.154..229H,2011AnPhy.326.3075H}.  These would
enter the hydrodynamical description of a dense quark-gluon plasma in the
cases where thermal and charge relaxations play a phenomenological
role. We discuss and update for completeness the shear viscosity of
quark matter which was already studied in
Refs.~\cite{2008JPhG...35c5003I,2008EPJA...38...97A,LW14,LKW15}.  We
consider the regime where the quark self-energies are dominated by 
mesonic fluctuations, \ie, the regime close to the chiral phase-transition line, 
which is relevant for heavy-ion collisions.  Although
our method and results can be straightforwardly applied to the dense
and cold regime of compact stars, a number of factors, such as 
nonzero isospin, color superconductivity, and the presence of leptons,
would require additional effort.

The paper is organized as follows. Section~\ref{sec:Kubo} derives 
the expressions for the thermal and electrical
conductivities  from the
Kubo-Zubarev formalism.  In Sec.~\ref{sec:QSpectralFunctions} we discuss the quark
and meson masses within the two-flavor NJL model and derive the
spectral function of quarks taking into account the Dirac structure of
the self-energies. Numerical results for the transport coefficients are
given in Sec.~\ref{sec:results}.  Our results are summarized in
Sec.~\ref{sec:conclusions}.  Appendix \ref{app:B} is devoted to the
equilibrium properties of the two-flavor NJL model, Appendix \ref{app:C} computes the quark self-energy due to meson exchange,
and finally Appendix \ref{app:D} lists some of the relevant
thermodynamic relations used in our computations.  We use natural
(Gaussian) units with $\hbar= c = k_B = k_e = 1$, $e=\sqrt{4\pi\alpha}$,
$\alpha=1/137$, and the metric signature $(+,-,-,-)$.

\section{Kubo formulas for transport coefficients}
\label{sec:Kubo}

The Kubo-Zubarev formalism relates the transport properties of a
statistical ensemble to different types of {\it equilibrium}
correlation functions, which in turn can be computed using equilibrium
many-body
techniques~\cite{1957JPSJ...12..570K,zubarev1997statistical}.  We
start our discussion with the Lagrangian of the underlying effective
model, as it will specify the power counting required for the
computation of the two-point correlation functions.

\subsection{Lagrangian}
\label{sec:NJL_Lagranian}

We consider two-flavor quark matter described by the NJL Lagrangian of
the form
\be\label{eq:lagrangian}
\mathcal{L}=\bar\psi(i\slashed \partial-m_0)\psi+
\frac{G}{2}\left[(\bar\psi\psi)^2+
(\bar\psi i\gamma_5\bm\tau\psi)^2\right],
\ee 
where $\psi=(u,d)^T$ is the isodoublet quark field, $m_0=5.5$ MeV is the
current quark mass, $G=10.1$ GeV$^{-2}$ is the effective four-fermion coupling
constant, and $\bm\tau$ is the vector of Pauli isospin matrices. This
Lagrangian describes four-fermion scalar-isoscalar and pseudoscalar-isovector
interactions between quarks with the corresponding bare vertices
$\Gamma^0_{s}=1$ and $\Gamma^0_{ps}=i\bm\tau\gamma_5$.  The
symmetrized energy-momentum tensor is given in the standard fashion by
\be\label{eq:energymom}
T_{\mu\nu}=\frac{i}{2}(\bar\psi\gamma_{\mu}
\partial_{\nu}\psi +\bar\psi\gamma_{\nu}
\partial_{\mu}\psi)-g_{\mu\nu}\mathcal{L},
\ee 
and the quark-number and charge currents are defined as
\be\label{eq:current}
N_{\mu}=\bar\psi\gamma_{\mu}\psi,\qquad
J_{\mu}=\bar\psi\hat Q\gamma_{\mu}\psi,
\ee 
where 
\be\label{eq:charge}
\hat Q=e 
\begin{pmatrix}
    2/3 & 0 \\
    0 & -1/3 
\end{pmatrix}
\ee 
is the charge matrix in flavor space, with $e$ being the elementary
charge. The expression for the energy-momentum tensor
\eqref{eq:energymom} is symmetric in its indices; this form is
necessary for the implementation in the Kubo formulas.

\subsection{Thermal and electrical conductivities}

Within the Kubo-Zubarev 
approach the thermal and electrical conductivities are given by
\bea\label{eq:kappa}
\kappa = -\frac{\beta}{3}\frac{d}{d\omega} {\rm Im}\Pi^{R}_\kappa (\omega)\Big|_{\omega =0},\\
\label{eq:sigma}
\sigma = -\frac{1}{3}\frac{d}{d\omega} {\rm Im}\Pi^{R}_\sigma (\omega)\Big|_{\omega =0},
\eea
where $\beta =T^{-1}$ is the inverse temperature, and the retarded
correlation functions on the right-hand sides are defined 
as~\cite{zubarev1997statistical,2011AnPhy.326.3075H}
\bea\label{eq:corkappasigma1}
\Pi^{R}_\kappa(\omega) &=& i\int_{0}^{\infty}dt\
 e^{i\omega t}\int d\bm r\langle
\left[q_{\mu}(\bm r,t),q^{\mu}(0)\right]\rangle_{0},\\
\label{eq:corkappasigma2}
\Pi^{R}_\sigma(\omega) &=& i\int_{0}^{\infty}dt\ 
e^{i\omega t}\int d\bm r\langle
\left[j_{\mu}(\bm r,t),j^{\mu}(0)\right]\rangle_{0},
\eea
\ie, they are the statistical averages of commutators 
(denoted by $[\cdot , \cdot ]$) of the heat and electrical currents defined,
respectively, as
\bea\label{eq:termcurrent1}
q_\mu &=& \Delta_{\mu\alpha}u_{\beta}T^{\alpha\beta}
-h\Delta_{\mu\alpha}N^{\alpha},\\
\label{eq:elcurrent1}
j_{\mu} &=&\Delta_{\mu\alpha}J^{\alpha}.
\eea
Here $u_{\beta}$ is the 4-velocity of the fluid,
$\Delta_{\mu\nu}=g_{\mu\nu}-u_{\mu}u_{\nu}$ is the projector on the
direction transverse to the fluid velocity, $h$ is the enthalpy per
particle, and the energy-momentum tensor $T^{\mu\nu}$ is assumed to be
symmetric in its indices.  Note that the heat current
\eqref{eq:termcurrent1} differs from the net energy flow by the
particle-convection term $\propto h$.

 Equations
\eqref{eq:kappa}-\eqref{eq:corkappasigma2} apply to arbitrary quantum
statistical ensembles without restrictions on the strength of the
couplings of the underlying theory. In the following we will derive
more specific expressions suitable for the NJL model with contact
scalar and pseudoscalar couplings among quarks by applying the
$1/N_c$ expansion to select the dominant diagrams contributing to the
correlation functions.

It is convenient to evaluate the correlation functions of interest
using the thermal equilibrium Green's functions of the imaginary-time
Matsubara technique. In the fluid rest frame $u_{\mu}=(1,0,0,0)$,
$\Delta_{\mu\nu}=\rm diag(0,-1,-1,-1)$, and the Matsubara correlation
functions read
\bea\label{eq:corkappa_m}
\Pi^{M}_\kappa(\omega_n) &=& \Pi^M_{TT}(\omega_n)-2h\Pi^M_{TN}(\omega_n)+h^2\Pi^M_{NN}(\omega_n),\nonumber\\
\\
\label{eq:corTT_m}
-\frac{1}{3}\Pi^{M}_{TT}(\omega_n)\!\!&=&\!\! \int_{0}^{\beta}d\tau 
e^{i\omega_n\tau}\!\!\int d\bm r\langle
T_{\tau}(T_{01}(\bm r,\tau),T_{01}(0))\rangle_0,\nonumber\\
\\
\label{eq:corTN_m}
-\frac{1}{3}\Pi^{M}_{TN}(\omega_n)\!\!&=&\!\! \int_{0}^{\beta}d\tau 
e^{i\omega_n\tau}\!\!\int d\bm r\langle
T_{\tau}(T_{01}(\bm r,\tau),N_{1}(0))\rangle_0,\nonumber\\
\\
\label{eq:corNN_m}
-\frac{1}{3}\Pi^{M}_{NN}(\omega_n)\!\!&=&\!\! \int_{0}^{\beta}d\tau 
e^{i\omega_n\tau}\!\!\int d\bm r\langle
T_{\tau}(N_{1}(\bm r,\tau),N_{1}(0))\rangle_0,\nonumber\\
\\
\label{eq:corsigma_m}
-\frac{1}{3}\Pi^{M}_\sigma(\omega_n)\!\!&=&\!\!
\int_{0}^{\beta}\!\!d\tau
 e^{i\omega_n\tau}\!\!\! \int \! d\bm r\langle
T_{\tau}(J_1(\bm r,\tau),J_1(0))\rangle_0,\nonumber\\
\eea
where $T_{01}(\bm r,\tau)$, $N_1(\bm r,\tau)$, and $J_1(\bm r,\tau)$
are obtained from $T_{01}(\bm r,t)$, $N_1(\bm r,t)$, and $J_1(\bm r,t)$
via Wick rotation $t\to -i\tau$, $T_\tau$ is the time-ordering
operator for imaginary time $\tau$, and the factor 3 arises from
summation over the directions of isotropic three-dimensional space.
In Eq.~\eqref{eq:corkappa_m} we decomposed the thermal current
according to Eq.~\eqref{eq:termcurrent1} and used the symmetry of the
correlation function with respect to its
arguments~\cite{2011AnPhy.326.3075H}.  The required retarded
correlation functions \eqref{eq:corkappasigma1} and
\eqref{eq:corkappasigma2} can be obtained from
(\ref{eq:corkappa_m})-(\ref{eq:corsigma_m}) by analytic continuation
$i\omega_n\to \omega +i\delta$.  Note that the transformation to
imaginary time implies a change of the derivative
$\partial_0\to i\partial_\tau$.  Because $T_{\mu\nu}$, and therefore
also $N_\mu$ and $J_\mu$, are bosonic operators, the Matsubara
frequencies assume even integer values $\omega_n=2\pi nT$,
$n=0,\pm 1,\ldots$.  The $T_{01}$ component of Eq.~(\ref{eq:energymom})
is given by
\be\label{eq:to1m}
T_{01}(\bm r,\tau)=i\bar\psi(\bm r,\tau)\frac{\gamma_0}{2}\partial_1\psi(\bm r,\tau) 
+i\bar\psi(\bm r,\tau)\frac{\gamma_1}{2}i\partial_\tau\psi(\bm r,\tau).\nonumber\\
\ee
Substituting $T_{01}(\bm r,\tau)$,
$N_1(\bm r,\tau)$, and $J_1(\bm r,\tau)$ into
Eqs.~(\ref{eq:corkappa_m})-(\ref{eq:corsigma_m}) we obtain
\bea\label{eq:corTTm1}
-\frac{1}{3}\Pi^{M}_{TT}(\omega_n)&=&\sum_{\alpha,\alpha'}\int_{0}^{\beta}
d\tau\ e^{i\omega_n\tau}\nonumber\\
&&\hspace{-1.9cm}\times\int d\bm r\langle
T_{\tau}(i\bar\psi\frac{\gamma_{\mu}}{2}
\partial_{\alpha}\psi\Big\vert_{(\bm r,\tau)},
i\bar\psi\frac{\gamma_{\mu'}}{2}\partial_{\alpha'}\psi\Big\vert_0)\rangle_0,\\
\label{eq:corTNm1}
-\frac{1}{3}\Pi^{M}_{TN}(\omega_n)&=&\sum_{\alpha}\int_{0}^{\beta}
d\tau\ e^{i\omega_n\tau}\nonumber\\
&&\hspace{-1.7cm}\times\int d\bm r\langle
T_{\tau}(i\bar\psi\frac{\gamma_{\mu}}{2}
\partial_\alpha\psi\Big\vert_{(\bm r,\tau)},\bar\psi \gamma_1\psi \Big\vert_0)\rangle_0,\\
\label{eq:corNNm1}
-\frac{1}{3}\Pi^{M}_{NN}(\omega_n)&=&\int_{0}^{\beta}d\tau\ 
e^{i\omega_n\tau}\nonumber\\
&&\hspace{-1.7cm}\times
\int d\bm r\langle T_{\tau}(\bar\psi \gamma_1\psi\Big\vert_{(\bm r,\tau)},
\bar\psi \gamma_1\psi\Big\vert_0)\rangle_0,\\
\label{eq:corsigmam1}
-\frac{1}{3}\Pi^{M}_\sigma(\omega_n)&=&\int_{0}^{\beta}d\tau\ 
e^{i\omega_n\tau}\nonumber\\
&&\hspace{-1.7cm}\times\int d\bm r\langle
T_{\tau}(\bar\psi \hat Q\gamma_1\psi\Big\vert_{(\bm r,\tau)},
\bar\psi \hat Q\gamma_1\psi\Big\vert_0)\rangle_0,
\eea
where $\alpha, \alpha', \mu, \mu'$ assume values $1, 0$, with
$\mu\neq \alpha, \mu'\neq \alpha'$, \ie, the sums in
Eqs.~\eqref{eq:corTTm1} and \eqref{eq:corTNm1} contain four and two
terms, respectively. 

\begin{figure}[tbh] 
\begin{center}
\includegraphics[width=7.5cm,keepaspectratio]{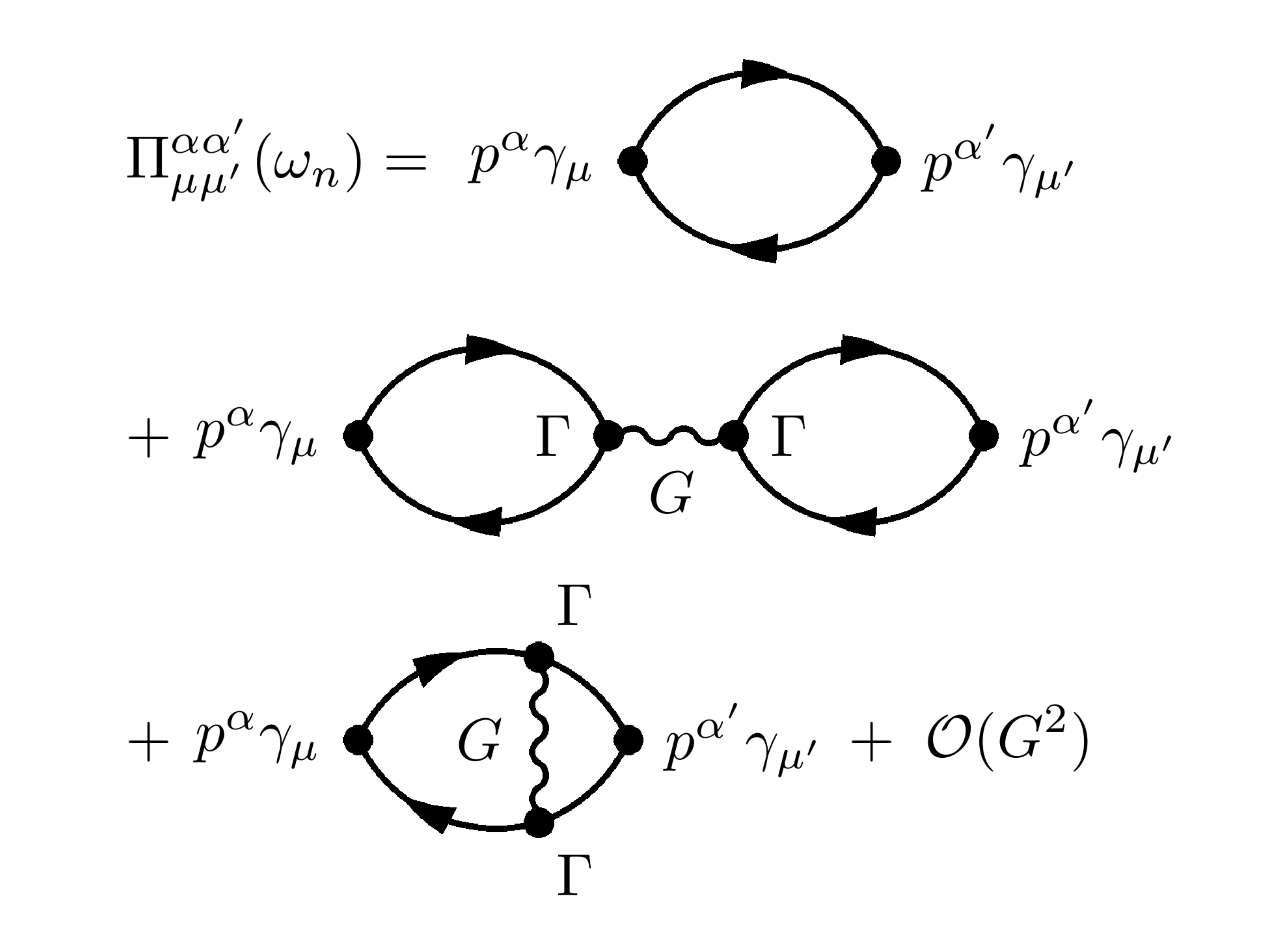}
\caption{ Contributions to the two-point correlation functions from 
  ${\cal O}(N_c^1)$ (first and second lines) and   ${\cal O}(N_c^0)$ (the 
third line) diagrams which contain a single interaction line $G$. 
}
\label{fig:loops} 
\end{center}
\end{figure}

To select the relevant diagrams contributing to the correlation
functions we apply the $1/N_c$ power-counting scheme, in which each
loop contributes a factor of $N_c$ from the trace over color
space~\cite{1994PhRvC..49.3283Q,2008JPhG...35c5003I,2008EPJA...38...97A,LW14,LKW15,2016PhRvC..93d5205G,LangDiss}.
Each coupling $G$ (which is associated with a pair of
$\Gamma^0_{s/ps}$ matrices) contributes a factor of
$1/N_c$. Therefore, for any given number of $\Gamma^0_{s/ps}$ vertices
the leading diagram in the $1/N_c$ approximation is the one that has the
maximum number of loops. Figure \ref{fig:loops} shows the diagrammatic
expansion for the two-point correlation function, which we define in a
generic form
\bea\label{eq:rings}
\Pi_{\mu\mu'}^{\alpha\alpha'}(\omega_n) &=&
\int_{0}^{\beta}\!\! d\tau e^{i\omega_n\tau}
\int d\bm r\nonumber\\
&\times &
\langle T_{\tau}(i\bar\psi\gamma_\mu 
\partial^\alpha\psi\Big\vert_{(\bm r,\tau)},
i\bar\psi\gamma_{\mu'}\partial^{\alpha'}\psi\Big\vert_0)\rangle_0,\quad
\eea 
where $\partial_\alpha=(i\partial_\tau,\partial_i)$. The diagrams are
arranged according to the $1/N_c$ expansion.  The first line contains
a loop and no coupling $G$ and is of order of $N_c$; the second line
contains two loops and a coupling, and therefore is again of order of
$N_c$; the third line, which contains a loop and a coupling, is of
order of $N_c^0$.
Thus, the correlation function \eqref{eq:rings} in the leading
[${\cal O}(N_c^1)$] order is given by a sum of loop diagrams
which contain the single-loop contribution (first line in
Fig.~\ref{fig:loops})
\bea\label{eq:ring1}
T\sum_l\!\! \int\!\! \frac{d\vecp}{(2\pi)^3}p^{\alpha}p^{\alpha'}
\Tr \left[
\gamma_{\mu} G(\vecp, i\omega_l+i\omega_n) \gamma_{\mu'} G(\vecp, i\omega_l) \right], \nonumber\\
\eea
plus multiloop contributions which necessarily contain loop
contributions of the type 
\bea\label{eq:ring2}
T\sum_l\!\! \int\!\! \frac{d\vecp}{(2\pi)^3}p^{\alpha}\Tr \left[
\gamma_{\mu} G(\vecp, i\omega_l+i\omega_n) \Gamma^0_{s/ps} G(\vecp, i\omega_l) \right]\!\!,\nonumber\\
\eea
see the second line in Fig.~\ref{fig:loops}.
Here $G(\vecp, i\omega_l)$ is the dressed Matsubara Green's function
of quarks, the summation goes over fermionic Matsubara frequencies
$\omega_l=\pi(2l+1)T-i\mu$, $l=0,\pm1,\ldots,$ with temperature $T$ and
chemical potential $\mu$; (in isospin-symmetric two-flavor quark
matter there is a single chemical potential for both $u$ and $d$
quarks). The traces should be taken in Dirac, color, and flavor
space. The Lorentz structure of the Green's function implies that 
(a)  diagrams of type \eqref{eq:ring2} with pseudoscalar vertices
vanish due to the trace over the Dirac space and (b) those with scalar
vertices vanish if $\alpha\neq \mu$, because the integrand has an odd
power of momentum, which implies that the momentum
integral vanishes in isotropic momentum space. Thus, the only term
contributing to Eq.~\eqref{eq:corTTm1} is the one-loop expression
\eqref{eq:ring1}. In the same way one can see that the multiloop
diagrams vanish also for the other three correlation functions
\eqref{eq:corTNm1}-\eqref{eq:corsigmam1}. Thus, for the correlation functions
\eqref{eq:corTTm1}-\eqref{eq:corsigmam1} we obtain
\bea\label{eq:feynmanrules_TT}
\frac{1}{3}\Pi^{M}_{TT}(\omega_n)&=&
\frac{T}{4}\sum\limits_l\sum_{\alpha,\alpha'}\int\!\! 
\frac{d\bm p}{(2\pi)^3}p_\alpha p_{\alpha'}\nonumber\\
&&\hspace{-1.4cm}\times
\Tr\bigl[\gamma_{\mu}
 G(\bm p, i\omega_l+i\omega_n)
\gamma_{\mu'} G(\bm p, i\omega_l)\bigr],\\
\label{eq:feynmanrules_TN}
\frac{1}{3}\Pi^{M}_{TN}(\omega_n)&=&
\frac{T}{2}\sum\limits_l\sum_{\alpha}\int \frac{d\bm p}{(2\pi)^3}p_\alpha \nonumber\\
&&\hspace{-1.4cm}\times
\Tr\bigl[\gamma_1 G(\bm p, i\omega_l+i\omega_n)
\gamma_{\mu} G(\bm p, i\omega_l)\bigr],\\
\label{eq:feynmanrules_NN}
\frac{1}{3}\Pi^{M}_{NN}(\omega_n)&=&
T\sum\limits_l\int \frac{d\bm p}{(2\pi)^3}
\nonumber\\
&&\hspace{-1.4cm}\times
\Tr\bigl[\gamma_1 G(\bm p, i\omega_l+i\omega_n)
\gamma_1 G(\bm p, i\omega_l)\bigr],\\
\label{eq:feynmanrules_sigma}
\frac{1}{3}\Pi^{M}_\sigma(\omega_n)&=&
T\sum\limits_l\int\!\! \frac{d\bm p}{(2\pi)^3}\nonumber\\
&&\hspace{-1.4cm}\times
\Tr\bigl[\hat Q\gamma_1 G(\bm p, i\omega_l+i\omega_n)
\hat Q\gamma_1 G(\bm p, i\omega_l)\bigr],
\eea
where $p_\alpha, p_{\alpha'}$ assume values $p_1$ and
$p_0=i\omega_l+i\omega_n/2$. Note that the expressions
\eqref{eq:feynmanrules_TT}-\eqref{eq:feynmanrules_sigma} are valid in
a wider context, \ie, in any relativistic theory (both bosonic and
fermionic) where the single-loop (skeleton) diagram with fully dressed
propagators constitutes the leading-order contribution in the power-counting scheme.

The Matsubara summations appearing in these
expressions can be cast into the general form
\bea\label{eq:sums}
S_{\mu\nu}\left[f\right](\bm p,
i\omega_n)&=&T\sum\limits_l 
\Tr\bigr[\gamma_\mu G(\bm p, i\omega_l+i\omega_n)
\nonumber\\
&\times & 
\gamma_\nu G(\bm p, i\omega_l)\bigr]
f(i\omega_l +i\omega_n/2),
\eea
where $f(z)=z^n$ with $n=0,1,2$. The summation is standard upon
introducing the spectral representation of the temperature Green's
functions
\be\label{eq:propagator}
G(\bm p, z)=\int_{-\infty}^{\infty}d\varepsilon 
\frac{A(\bm p, \varepsilon)}{z-\varepsilon},
\ee 
where the spectral function is given by  
\be\label{eq:spectralfunction0}
A(\bm p, \varepsilon)=-\frac{1}{2\pi i}\left[G^R(\bm p, \varepsilon)-
G^A(\bm p, \varepsilon)\right],
\ee
and $G^{R/A}$ are the retarded/advanced Green's functions. After
summation and subsequent analytical continuation 
$i\omega_n\to \omega +i\delta$ we find
\bea\label{eq:residue4}
S_{\mu\nu}[f](\bm p, \omega)&=&\int_
{-\infty}^{\infty} \!\! d\varepsilon  \!\!\int_
{-\infty}^{\infty}\!\! d\varepsilon' 
\Tr[\gamma_\mu A(\bm p, \varepsilon') 
\gamma_\nu A(\bm p, \varepsilon)]
\nonumber\\
&\times&
\frac{\tilde n(\varepsilon)f(\varepsilon+\omega/2)
-\tilde n(\varepsilon')f(\varepsilon'-\omega/2)}
{\varepsilon-\varepsilon' +\omega+i\delta},\qquad
\eea
where $n(\varepsilon)=[e^{\beta(\varepsilon-\mu)}+1]^{-1}$ is the
Fermi distribution function and
$\tilde n(\varepsilon) = n(\varepsilon)-1/2.$ Substituting this result
into the correlation functions
\eqref{eq:feynmanrules_TT}-\eqref{eq:feynmanrules_sigma} we obtain
compact expressions in terms of Eq.\ \eqref{eq:residue4}
\bea\label{eq:feynmanrules1}
\frac{1}{3}\Pi_{TT}(\omega)&=&
\frac{1}{4}\int \frac{d\bm p}{(2\pi)^3}\big\{p_1^2
S_{00}[f=1]\nonumber\\
&&\hspace{-0.9cm}+2p_1S_{01}[f=\varepsilon]+
S_{11}[f=\varepsilon^2]\big\}(\bm p, \omega),\\
\label{eq:feynmanrules2}
\frac{1}{3}\Pi_{TN}(\omega)&=&
\frac{1}{2}\int \frac{d\bm p}{(2\pi)^3}
\big\{p_1 S_{10}[f=1]\nonumber\\
&+&S_{11}[f=\varepsilon]\big\}(\bm p, \omega),\\
\label{eq:feynmanrules3}
\frac{1}{3}\Pi_{NN}(\omega)&=&
\int \frac{d\bm p}{(2\pi)^3}
S_{11}[f=1](\bm p, \omega),\\
\label{eq:feynmanrules4}
\frac{1}{3}\Pi_\sigma(\omega)&=&
\frac{1}{3}\Pi_{NN}(\omega)\times 
\frac{\Tr\hat Q^2}{N_f},
\eea
with $N_f$ being the number of flavors. These expressions can now be
substituted into Eqs.~\eqref{eq:kappa} and \eqref{eq:sigma} to find the
thermal and electrical conductivities. It is, however, convenient to first
separate the real and imaginary parts in Eq.~\eqref{eq:residue4} via
the Dirac identity in order to find the required $\omega\to 0$
limit. We find from
Eq.~\eqref{eq:residue4}
\bea\label{eq:difimsums}
\frac{d}{d\omega}{\rm Im} S_{\mu\nu}[f]
(\bm p, \omega)\bigg|_{\omega=0}\!\!=
\pi\int_{-\infty}^{\infty}\!\! d\varepsilon
\frac{\partial n(\varepsilon)}{\partial \varepsilon } f(\varepsilon)
{\cal T}_{\mu\nu}(\bm p, \varepsilon),\nonumber\\
\eea
where 
\be\label{eq:TraceA2}
{\cal T}_{\mu\nu}(\bm p, \varepsilon)\equiv \Tr[
\gamma_\mu A(\bm p, \varepsilon)
\gamma_\nu A(\bm p, \varepsilon)].
\ee
Using expressions \eqref{eq:feynmanrules1}-\eqref{eq:TraceA2} we
finally obtain from Eqs.~\eqref{eq:kappa} and \eqref{eq:sigma}
\bea\label{eq:kappa2}
\kappa &=&-\frac{\pi}{4T}\int_{-\infty}^{\infty}
d\varepsilon \frac{\partial n(\varepsilon)}{\partial \varepsilon } \int \frac{d\bm p}
{(2\pi)^3}\bigl[p_1^2{\cal T}_{00}(\bm p,\varepsilon)\nonumber\\
&+&
2p_1(\varepsilon -2h){\cal T}_{01}(\bm p,\varepsilon)+
(\varepsilon -2h)^2 {\cal T}_{11}(\bm p,\varepsilon)\bigr],
\\
\label{eq:sigma2}
\sigma&=&-\pi\Tr \frac{\Tr\hat Q^2}{N_f}\int_{-\infty}^{\infty} 
d\varepsilon\frac{\partial n(\varepsilon)}{\partial \varepsilon }  \int \frac{d\bm p}
{(2\pi)^3}{\cal T}_{11}(\bm p,\varepsilon).
\eea
Thus, the problem of computing the transport coefficients reduces to
the determination of the spectral function of the quarks followed by
computing the components of the trace
${\cal T}_{\mu\nu}(\bm p, \varepsilon)$.

The quark spectral function in an isotropic medium has a
general decomposition in terms of Lorentz-invariant components
\bea\label{eq:spectral}
A(\bm p, p_0)=-\frac{1}{\pi}(mA_s+p_0\gamma_0 A_0-\bm p \cdot \bm\gamma A_v),
\eea
where the coefficients $A_s, A_0, A_v$ are expressed in terms of the
analogous components of the self-energy in Appendix \ref{app:B}.
Substituting the decomposition  \eqref{eq:spectral} into Eqs.~\eqref{eq:kappa2}
and \eqref{eq:sigma2} we obtain
\bea\label{eq:kappa3}
\kappa =&-&\frac{N_cN_f}{\pi T}\int_{-\infty}^{\infty} \!\!
d\varepsilon  \frac{\partial n(\varepsilon) }{\partial \varepsilon} \int \!\!
\frac{d\bm p}{(2\pi)^3}
\Big\{p_1^2(A_s^2m^2+A_0^2\varepsilon ^2\nonumber\\
&+&A_v^2\bm p^2)+4p_1^2\varepsilon (\varepsilon -2h) A_0A_v-(\varepsilon -2h)^2\nonumber\\
&\times &(A_s^2m^2-A_0^2\varepsilon ^2+
A_v^2\bm p^2-2A_v^2p_1^2)\Big\},\\
\label{eq:sigma3}
\sigma =&-&\frac{4N_c}{\pi}\Tr \hat Q^2
\int_{-\infty}^{\infty} \!\!d\varepsilon
\frac{\partial n(\varepsilon) }{\partial \varepsilon}\int \!\!\frac{d\bm p}{(2\pi)^3}\nonumber\\
&\times & (-A_s^2m^2+A_0^2
\varepsilon ^2-A_v^2\bm p^2+2A_v^2p_1^2),
\eea
where we summed over the quark flavor ($N_f$)
and color ($N_c$) numbers.
Finally we note that the Lorentz-invariant
coefficients $A_s, A_0, A_v$  of the decomposition of the spectral
function depend only on
$\bm p^2$ and $\varepsilon$ (see Appendix \ref{app:B}), therefore the
angular integration can be done trivially by substituting
$p_1^2\to\bm p^2/3\equiv p^2/3$, after which we finally obtain 
\bea\label{eq:kappafinal}
\kappa &=&-\frac{N_cN_f}{6\pi^3T} \int_{-\infty}^{\infty}\!\!
d\varepsilon \frac{\partial n}{\partial \varepsilon}\int_{0}^{\Lambda} dp {p^2}
\Big\{\big[A_s^2(p,\varepsilon)m^2\nonumber\\
&-&A_0^2(p,\varepsilon)\varepsilon ^2
+A_v^2(p,\varepsilon)p^2\big]
[p^2-3(\varepsilon -2h)^2]\nonumber\\
&+& 2\big[A_0(p,\varepsilon)\varepsilon
+A_v(p,\varepsilon)(\varepsilon -2h)\big]^2p^2\Big\},\\
\label{eq:sigmafinal}
\sigma &=&\frac{40N_c\alpha}{27\pi^2}\int_{-\infty}^{\infty} \!\!
d\varepsilon \frac{\partial n}{\partial \varepsilon}\int_{0}^{\Lambda} dp p^2
\big[3A_s^2(p,\varepsilon)m^2\nonumber\\
&-&3A_0^2(p,\varepsilon)
\varepsilon ^2+A_v^2(p,\varepsilon)p^2\big],
\eea
where $\Lambda = 650$ MeV is the ultraviolet cutoff of the NJL model.
Given the Lorentz components of the spectral function we are in a
position to compute the thermal and electrical conductivities of
two-flavor quark matter using our final expressions 
\eqref{eq:kappafinal} and \eqref{eq:sigmafinal}.

\subsection{Shear viscosity}

Within the Kubo formalism the shear viscosity is given as \cite{1984AnPhy.154..229H}
\be\label{eq:eta}
\eta=-\frac{1}{10}\frac{d}{d\omega}\rm {Im}\Pi^R_\eta(\omega)\bigg|_{\omega=0},
\ee
where the retarded correlation function has the form
\be\label{eq:corpi1}
\Pi^{R}_\eta(\omega) = -i\int_{0}^{\infty}dt\ e^{i\omega t}\int d\bm r\langle
\left[\pi_{\mu\nu}(\bm r,t),\pi^{\mu\nu}(0)\right]\rangle_{0},
\ee
with $\pi_{\mu\nu}$ being the shear-viscosity tensor, defined as
\bea\label{eq:shear1}
&&\pi_{\mu\nu}=\Delta_{\mu\nu}^{\alpha\beta}T_{\alpha\beta},
\eea
where
\be\label{eq:delta}
\Delta_{\mu\nu}^{\alpha\beta}=\frac{\Delta_{\mu}^{\alpha}\Delta_{\nu}^\beta
+\Delta_{\mu}^{\beta}\Delta_{\nu}^\alpha}{2}-\frac{1}{3}\Delta_{\mu\nu}\Delta^{\alpha\beta}.
\ee
It is useful to note that
$\Delta_{\mu\nu}^{\alpha\beta}g_{\alpha\beta}=0$ by definition,
therefore the component of the energy-momentum tensor
\eqref{eq:energymom} containing $g_{\mu\nu}$ does not contribute to
Eq.~\eqref{eq:shear1}.

In the fluid rest frame $\Delta_{i}^{j}=\delta_{ij}$,
$\Delta_{0}^{0}=\Delta_{0}^{j}=0$, where $i,j=1,2,3$, and
$\delta_{ij}$ is the Kronecker symbol. In this frame only the spatial
components of Eq.~\eqref{eq:shear1},
$\pi_{ij}=T_{ij}-\delta_{ij}T_{mm}/3$, are nonzero.  Then, the
two-point correlation function  (\ref{eq:corpi1}) takes the form
\bea\label{eq:corpi2}
\Pi^{R}_\eta(\omega) 
&=&-2i\!\int_{0}^{\infty}\!\!\!\! dt\!\!\ e^{i\omega t}
\!\!\int\!\! d\bm r\langle [T_{11},T_{11}]\nonumber\\
&&\hspace{1cm}-[T_{11},T_{22}]+3[T_{12},T_{12}]\rangle_{0},
\eea
where we took into account the isotropy of the medium and for the sake of brevity
omitted the arguments of the $T_{ij}$ tensors. Note that the commutator
$[T_{11},(T_{11}-T_{22})]$ is manifestly nonzero at the operator
level, and a computation shows that its statistical average does not
vanish in an isotropic medium and cannot be neglected.  The Matsubara
counterpart of this retarded two-point function is given by
\bea\label{eq:corpim1}
\Pi^M_\eta(\omega_n)&=&-2\!\int_{0}^{\beta}\!d\tau
e^{i\omega_n \tau}\!\int\! d\bm r\langle
T_\tau(T_{11}(\bm r,\tau)T_{11}(0)\nonumber\\
&&\hspace{-0.5cm}-
T_{11}(\bm r,\tau)T_{22}(0)+3T_{12}(\bm r,\tau)T_{12}(0))\rangle_{0}. 
\eea
We next compute the components of the energy-momentum tensor
contributing to Eq.~(\ref{eq:corpim1}), which leads to the expression
\bea\label{eq:corpim2} 
\Pi^M_\eta(\omega_n)&=&-2\Pi_{11}^{xx}+2\Pi_{12}^{xy}\nonumber\\
&-&
\frac{3}{2}(\Pi_{11}^{yy}
+\Pi_{12}^{yx}+\Pi_{21}^{xy}+\Pi_{22}^{xx}),
\eea
where the lower indices indicate the components of the Dirac matrices,
whereas the upper indices indicate the component of the
spatial derivative, see Eq.~\eqref{eq:rings}.
The last four terms in Eq.~\eqref{eq:corpim2} obtain contributions
only from one-loop diagrams by the same arguments as before, see
Eqs.~\eqref{eq:ring1} and \eqref{eq:ring2}. For the first two terms in
Eq.~\eqref{eq:corpim2} the diagrams containing more than one loop do
not vanish. However, the multiloop contributions cancel each other
after integration due to the isotropy. Thus we conclude that only
one-loop diagrams are contributing to Eq.~\eqref{eq:corpim2}.
After carrying out the Matsubara sums and analytical continuation we
obtain the retarded correlator, which we write using
Eq.~\eqref{eq:residue4} as
\bea\label{eq:feynmanrulespi1}
\Pi^R_\eta(\omega) &=& 2\int\!\! \frac{d\bm p}{(2\pi)^3}\bigg[p_x^2S_{11}-p_xp_yS_{21}+
\frac{3}{4}\big(p_y^2S_{11}\nonumber\\
&+&p_xp_yS_{21}+p_xp_yS_{12}+p_x^2S_{22}\big)\bigg],
\eea
where we have suppressed the $(\bm p, \omega)$ arguments of the
$S_{\mu\nu}[f=1]$ functions. We obtain the shear viscosity 
from the Kubo-Zubarev formula \eqref{eq:eta}
by using the relation \eqref{eq:difimsums} and the symmetry
$p_x\leftrightarrow p_y$ 
\bea\label{eq:eta1}
\eta& =& -\frac{\pi}{10}\int_{-\infty}^{\infty} \!\!
d\varepsilon \frac{\partial n(\varepsilon)}{\partial \varepsilon}
\!\!\int \!\!
\frac{d\bm p}{(2\pi)^3}\left[(2p_x^2+3p_y^2)
{\cal T}_{11} \right. \nonumber \\
&   & \hspace*{4cm} \left. +p_xp_y{\cal T}_{12}\right],
\eea
where the ${\cal T}_{\mu\nu}$ tensor is defined in Eq.~\eqref{eq:TraceA2}.
Substituting the decomposition for the spectral
function \eqref{eq:spectral}  into this expression we obtain
\bea\label{eq:eta2}
\eta 
&=&-\frac{2N_cN_f}{5\pi}\int_{-\infty}^{\infty} 
d\varepsilon \frac{\partial n(\varepsilon)}{\partial \varepsilon}\int_0^\Lambda
\frac{dp}{2\pi^2}p^2\int 
\frac{d\Omega}{4\pi}\nonumber\\
&&\big[5p_x^2(-A_s^2m^2+A_0^2\varepsilon ^2
-A_v^2p^2)+4A_v^2p_x^4+8A_v^2p_x^2p_y^2\big]\nonumber\\
&=&\frac{N_cN_f}{15\pi^3 }\int_{-\infty}^{\infty} 
d\varepsilon \frac{\partial n}{\partial \varepsilon}
\int_{0}^{\Lambda} dp\, p^4\nonumber\\
&&
\big[5A_s^2(p,\varepsilon)m^2-5A_0^2(p,\varepsilon)
\varepsilon ^2+A_v^2(p,\varepsilon)p^2\big].
\eea
We conclude that, as in the case of the thermal and electrical
conductivities, the knowledge of the Lorentz components of the
spectral function completely determines the shear viscosity of quark
matter.  We note that the expression~\eqref{eq:corpi2} for the
correlation function is consistent with the expressions given in
Refs.~\cite{2008JPhG...35c5003I, 2008EPJA...38...97A,LW14}, in
which only the $[T_{12},T_{12}]$ commutator appears, because the
spatial isotropy implies the relation
$[T_{11},T_{11}]=[T_{11},T_{22}]+2[T_{12},T_{12}]$.

\section{Quark spectral function in the two-flavor NJL model}
\label{sec:QSpectralFunctions}

Equations \eqref{eq:kappafinal}, \eqref{eq:sigmafinal}, and
\eqref{eq:eta2} provide the general expressions for the transport
coefficients in terms of the Lorentz components of the quark spectral function. 
Further progress requires the knowledge of the specific form
of this spectral function in the regime of physical interest, which is
determined by the elementary processes that lead to a nonzero imaginary
part of the quark self-energy. We turn now to the derivation of
these components within the two-flavor NJL model and start with the
phase structure of matter predicted by this model.

\begin{figure}[t] 
\begin{center}
\includegraphics[height=2cm]{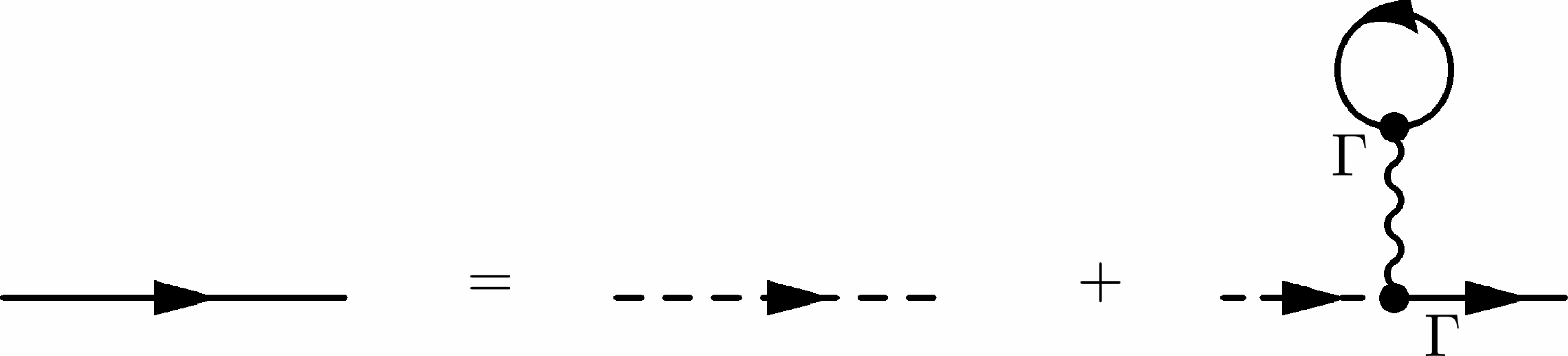}
\caption{ Dyson-Schwinger equation for the constituent quark mass.
  The dashed and solid lines are the bare and dressed propagators,
  respectively, and $\Gamma=1$. }
\label{fig:Hartree} 
\end{center}
\end{figure}

\begin{figure}[b] 
\begin{center}
\includegraphics[width=8.0cm,keepaspectratio]{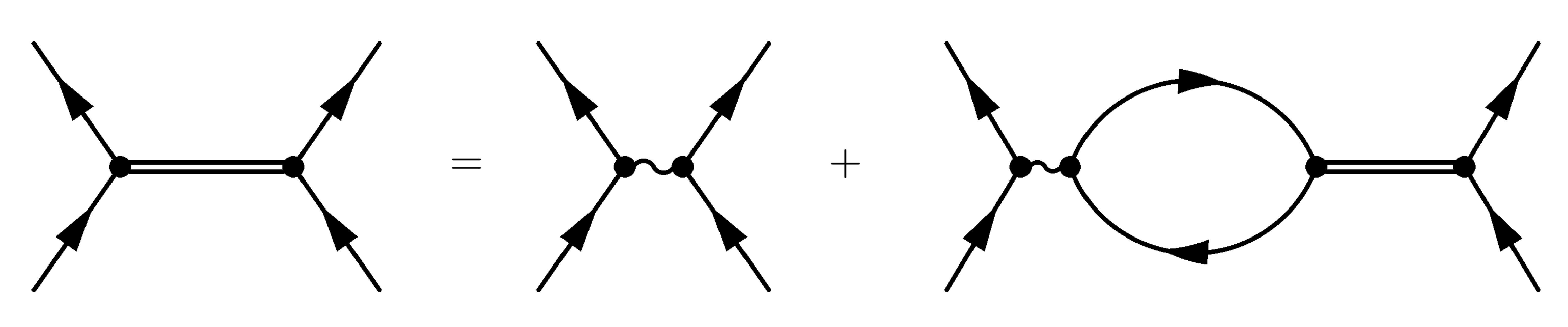}
\caption{ Bethe-Salpeter equation for mesons: 
 the double lines are the dressed meson propagators. 
}
\label{fig:BS_eq} 
\end{center}
\end{figure}

\begin{figure}[t] 
\begin{center}
\includegraphics[width=8.8cm,keepaspectratio]{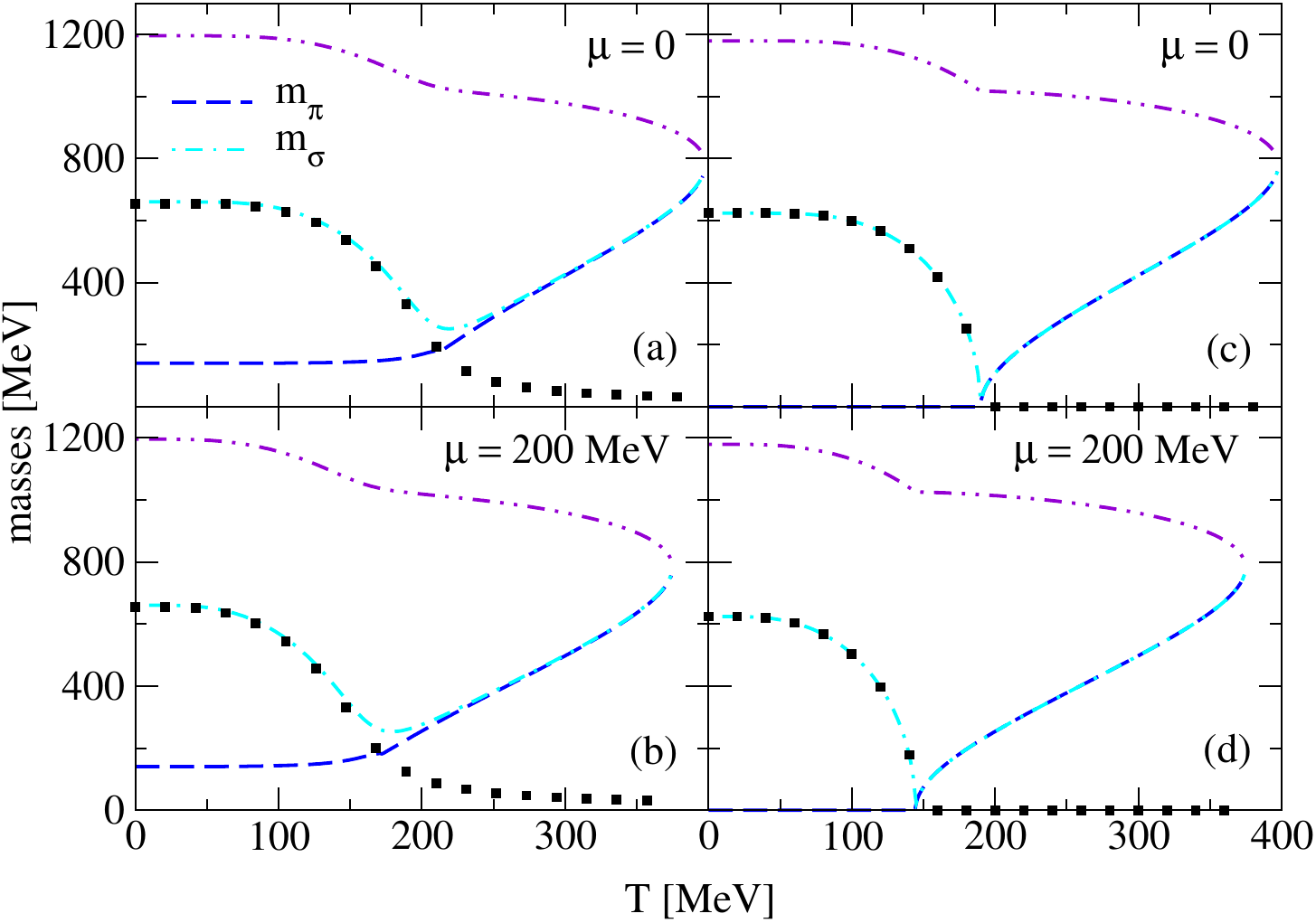}
\caption{ Quark and meson masses as functions 
of temperature at chemical potential $\mu = 0$ (upper row) and $\mu = 200$ MeV (lower row). 
The symbols correspond to twice the quark mass, the dashed and dash-dotted
lines correspond to the low-mass mesonic solutions, while the
dash-double-dotted lines correspond to 
the high-mass mesonic solution (see discussion in text).
Left panels: $m_0 > 0 $, right panels: $m_0=0$. }
\label{fig:masses} 
\end{center}
\end{figure}

At nonzero temperature and density the constituent quark mass
$m(T,\mu)$ is found to leading $\mathcal{O}(N_c^0)$ order in the $1/N_c$
expansion from a Dyson-Schwinger equation, where the self-energy is taken in
the Hartree approximation, see Fig.~\ref{fig:Hartree}.  The mesonic
propagator is obtained from the Bethe-Salpeter equation, shown in
Fig.~\ref{fig:BS_eq}, which resums contributions from quark-antiquark
polarization insertions. The meson masses are obtained as the poles of
the propagator in real space-time for $\bm p=0$.
The relevant calculations are reviewed in Appendix \ref{app:B}.  The
behavior of quark and meson masses as functions of density and
temperature are shown in Fig.~\ref{fig:masses} in the cases of explicitly broken
chiral symmetry $m_0\neq 0$ as well as the chiral limit $m_0 =0$.

As seen from Fig.~\ref{fig:masses}, there is always a nontrivial
solution for the quark masses with $m>m_0$ if chiral symmetry is
explicitly broken.  In the chiral limit, for fixed chemical potential,
the quark mass is nonzero below a certain temperature $T\le
T_{c}=T_{\rm M0}$
and is strictly zero for $T\ge T_{c}$.  More generally, at
sufficiently high densities and temperatures (for example,
$T>T_c\simeq 190$ MeV for $\mu=0$ or $\mu>\mu_c\simeq 330$ MeV for
$T=0$) one finds that chiral symmetry is restored ($m_0=m=0$).

The meson masses found from the Bethe-Salpeter equation for the meson
propagator are also shown in Fig.~\ref{fig:masses}. At sufficiently low 
temperatures and densities we find two solutions for the masses of the
$\pi$ and $\sigma$ mesons. The two low-mass solutions correspond to
the masses of the well-known $\pi$ and $\sigma$ mesons, and they satisfy
numerically the relation $m_\sigma^2=m_\pi^2+4m^2$ within 2\% precision.
The high-mass solutions are approximately the same for the scalar and
pseudoscalar modes and may correspond to a resonance state.
Note that in the chiral limit $m_0=0$ the low-mass solutions are given
by $m_\pi=0$ and $m_\sigma=2m$ below the critical temperature $T_c$ for
chiral phase transition. Above the critical temperature these
solutions become degenerate. As seen from Fig.~\ref{fig:masses}, the
lower and upper solutions approach each other with increasing
temperature and coincide at a temperature $T_{\rm max}\simeq 400$ MeV in
the case $\mu=0$. This limiting temperature decreases with
increasing chemical potential.

\begin{figure}[t] 
\begin{center}
\includegraphics[width=7.5cm,keepaspectratio]{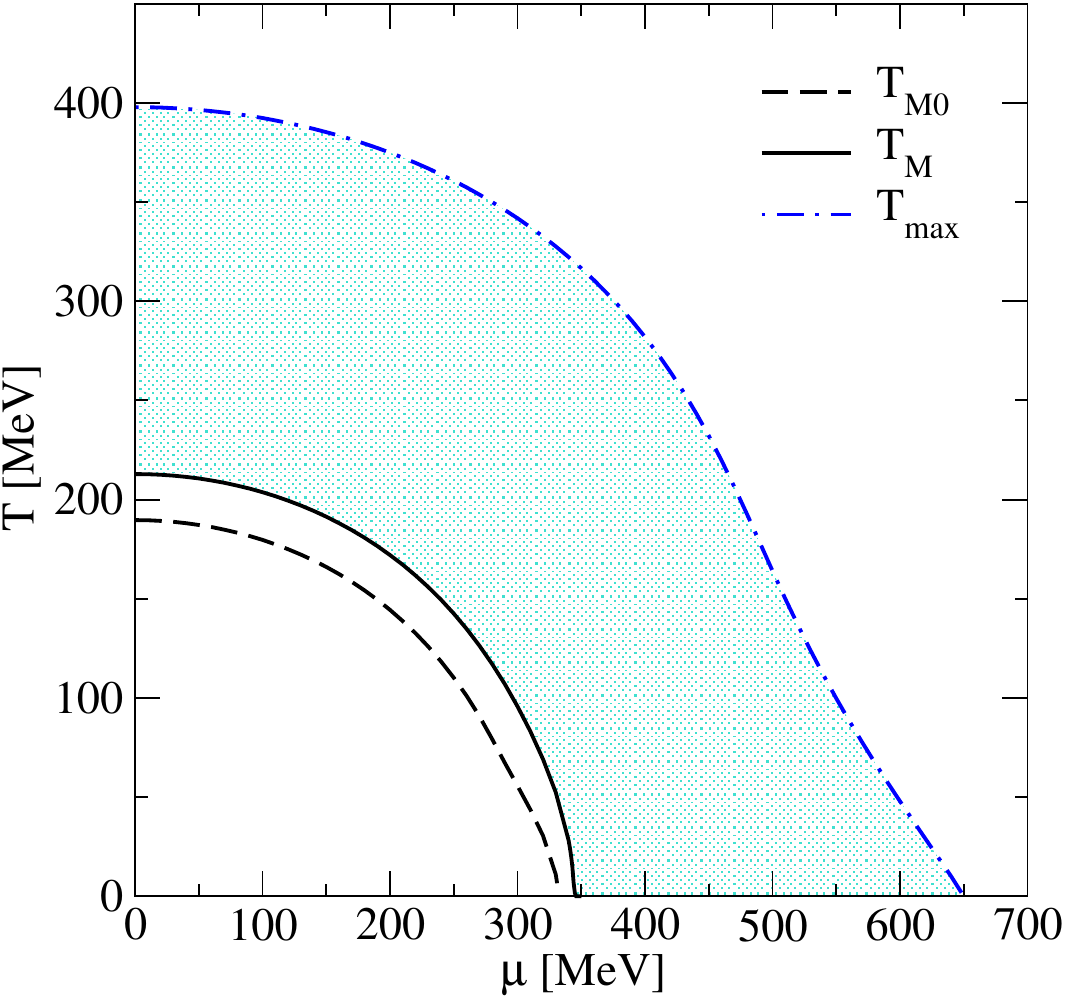}
\caption{ The Mott temperature $T_{\rm M}$ and the temperature $T_{\rm max}$
  (see discussion in the text) as functions of the chemical
  potential. The dashed line is the Mott temperature in the chiral
  limit $T_{\rm M 0}\equiv T_c$. The shaded area shows the portion of the phase diagram where
  our computations are applicable.}
\label{fig:mott_temp1} 
\end{center}
\end{figure}

Above $T_{\rm max}$ no solutions are found for the meson masses anymore, \ie,
the mesonic modes exist only for $T\le T_{\rm max}$ within
the zero-momentum pole approximation for the meson propagator. The maximal
temperature of existence of mesons $T_{\rm max}$ versus chemical
potential is shown in Fig.~\ref{fig:mott_temp1}.
In the limit $T\to 0$ the transition line ends
at $\mu_{\rm max}=\Lambda$, which implies $m_M\to 2\Lambda$. 

Another important temperature shown here is the Mott temperature
$T_{\rm M}$, which is defined by the condition $m_\pi=2m$ in the cases
$m_0=0$ and $m_0\neq 0$.  Above this temperature $m_\pi>2m$, and the
pion can decay into an on-shell quark-antiquark pair. As seen from
Figs.~\ref{fig:masses} and \ref{fig:mott_temp1}, $T_{\rm M}$ decreases
with chemical potential from the value $T_{\rm M} \simeq 213$ MeV at
$\mu=0$ and vanishes at $\mu\approx 345$ MeV. It coincides with
the chiral transition temperature in the chiral limit $m_0=0$.

To express the components of the spectral function in terms of the
self-energies we write the full quark retarded/advanced Green's
function as
\be\label{eq:propagator1} 
G^{R/A}(p_0,\bm p)=\frac{1}{\slashed p-m-\Sigma^{R/A}(p_0,\bm p)},
\ee
where $\Sigma^{R/A}$ is the quark-antiquark retarded/advanced
self-energy, which in the most general case (due to parity
conservation, translational and rotational invariance, as well as
time-reversal invariance) can be written in the following form
\bea\label{eq:selfenergy}
\Sigma^{R(A)}
=m\Sigma_s^{(*)}
-p_0\gamma_0\Sigma_0^{(*)}
+\bm p \cdot \bm\gamma\Sigma_v^{(*)}.
\eea
According to the definition of the spectral function
\eqref{eq:spectralfunction0} we find
\bea\label{eq:spectralfunction}
A(p_0,\bm p)
&=&-\frac{1}{\pi}(mA_s+p_0\gamma_0 A_0
-\bm p \cdot \bm\gamma A_v),
\eea
with
\bea\label{eq:spectral_coeff}
&&A_i=\frac{1}{d} 
[n_1\varrho_i -2n_2 (1+r_i) ],\quad 
d=n_1^2+4n_2^2,
\eea
where 
\bea
\label{eq:N1}
n_1&=&p_0^2[(1+r_0)^2-\varrho_0^2]\nonumber\\
&-&\bm p^2[(1+r_v)^2-\varrho_v^2]
-m^2[(1+r_s)^2-\varrho_s^2],\\
\label{eq:N2}
n_2&=&p_0^2\varrho_0 (1+r_0) \nonumber\\
&-&\bm p^2\varrho_v(1+r_v)-m^2\varrho_s(1+r_s),
\eea
with the short-hand notations $\varrho_i = {\rm Im}\Sigma_i$ and
$r_i = {\rm Re}\Sigma_i$, $i=s,0,v$.  From now on we will neglect the irrelevant real parts of the self-energy, 
which lead to momentum-dependent corrections to the constituent quark mass in next-to-leading
order ${\cal O} (N_c^{-1})$. 

\begin{figure*}[!] 
\begin{center}
\includegraphics[width=14.0cm,keepaspectratio]{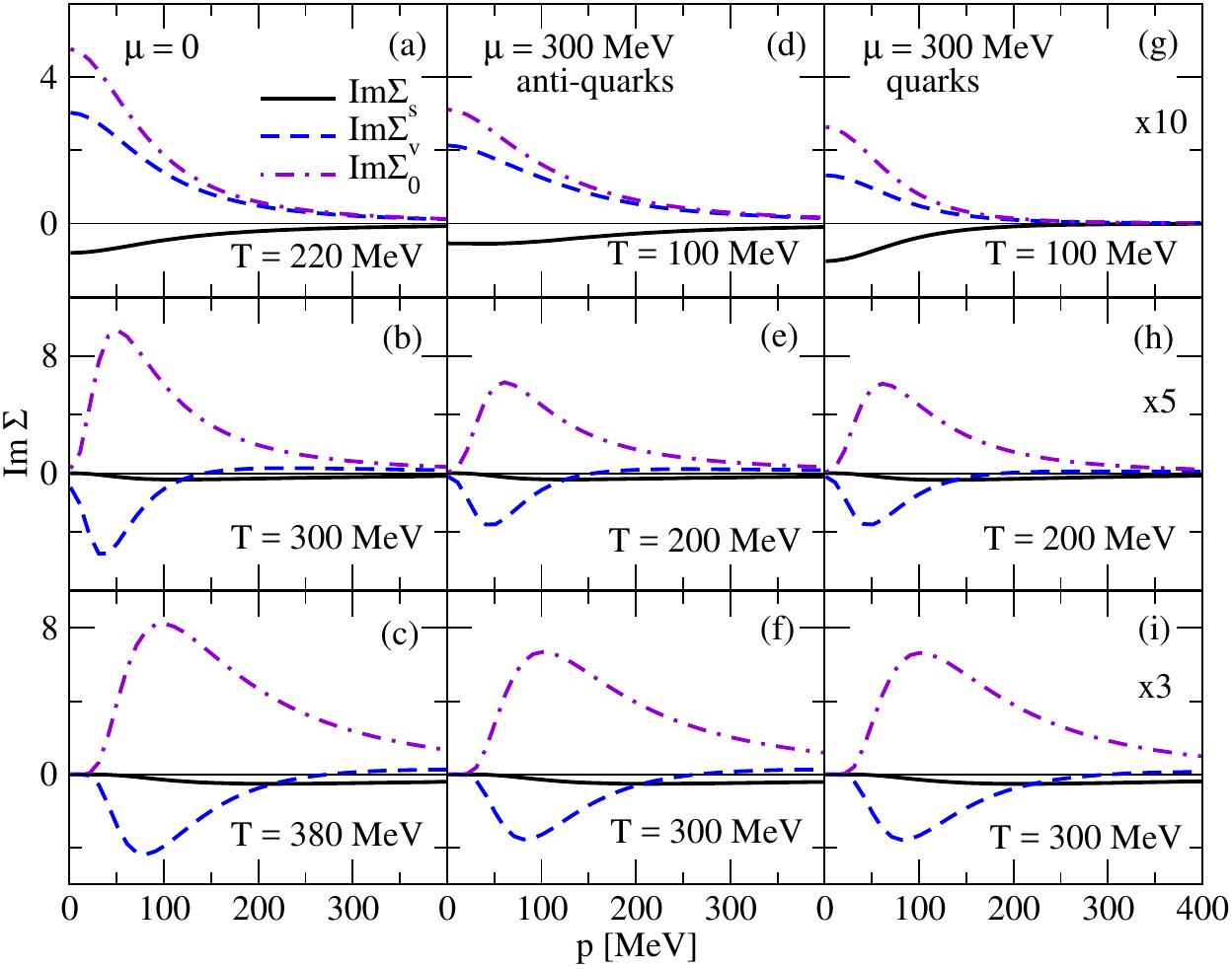}
\caption{ The imaginary parts of the three Lorentz components of the quark and antiquark on-shell self-energies as 
  functions of momentum at various values of temperature and chemical potential. 
  The signs of the antiquark self-energies have been inverted.}
\label{fig:self_energy} 
\end{center}
\end{figure*}

We now consider the quark self-energy that contributes to the
transport phenomena because of a nonvanishing imaginary part, closely
following similar computations by
Refs.~\cite{LKW15,LangDiss}.  The dominant
processes, according to the discussion of the phase structure above,
are the meson decays into two quarks and the inverse process above the
Mott temperature $T_{\rm M}$. The quark self-energy arising from meson
exchange is given in Matsubara space by
\bea
  \Sigma^M(\bm p,\omega_n) &=&
 \begin{minipage}{3cm}
\includegraphics[width=1.\textwidth]{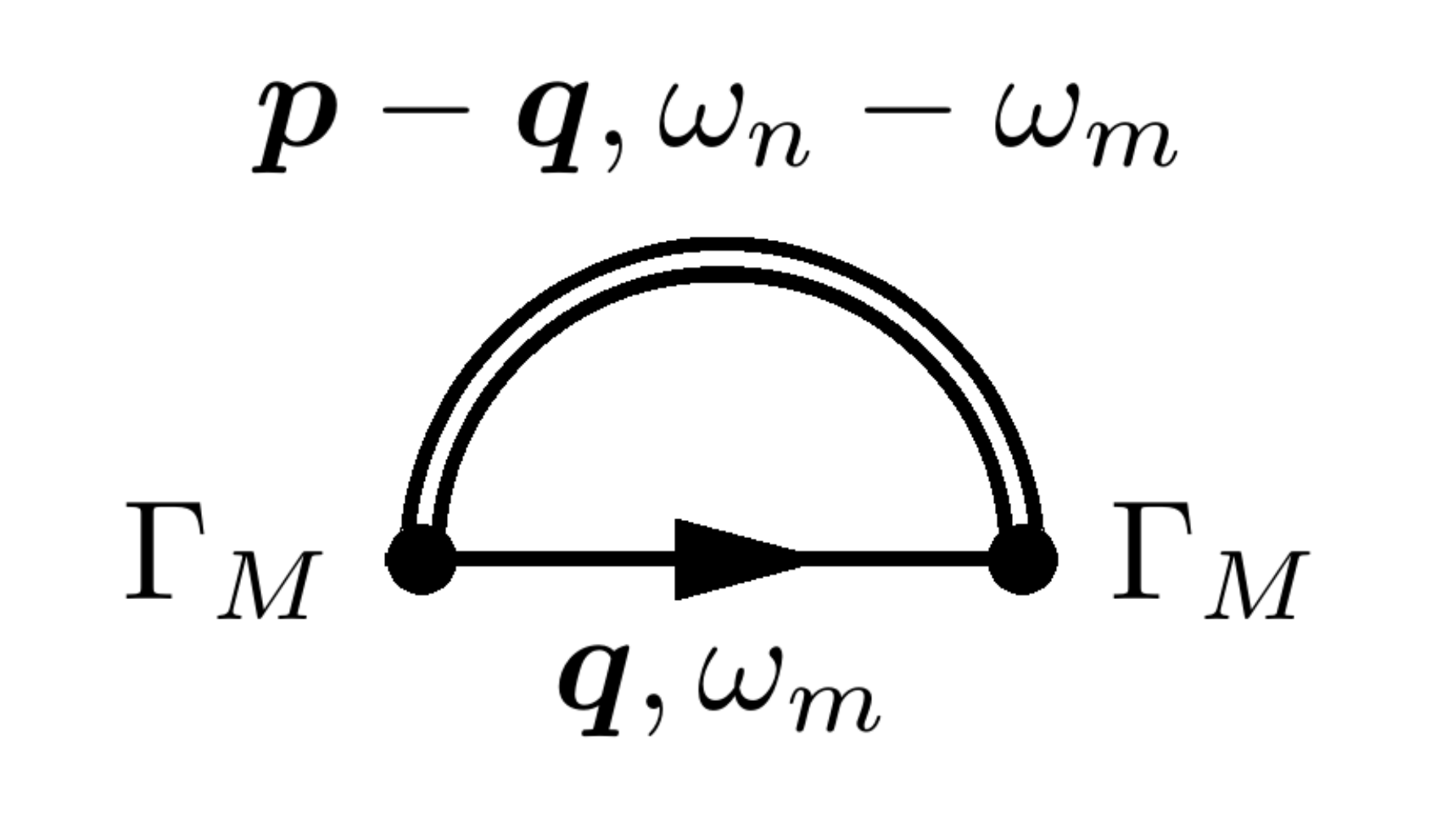} 
\end{minipage}
\nonumber\\
&&\hspace{-2.3cm} =T \sum_m \int \frac{d{\bm q}}{(2\pi)^3} 
\left[\Gamma_M S(\bm q, \omega_m) 
\Gamma_M D_M(\bm p-\bm
  q,\omega_n-\omega_m) \right],\nonumber\\
\label{eq:self}
\eea
where $S(\bm q, \omega_m)$ is the quark propagator with constituent
mass, and the index $M=\pi, \sigma$ indicates the meson. Using
$\Gamma_\sigma=1$ and $\Gamma_\pi=i\gamma_5\bm\tau$ we find the
decomposition
\bea\label{eq:self1}
\Sigma^M(\bm p,\omega_n)=
P_Mm\Sigma^M_s+i\omega_n\gamma_0\Sigma^M_0-\bm p \cdot \bm \gamma\Sigma^M_v, 
\eea
where $P_\sigma=1$, $P_\pi=-1$ and 
\bea\label{eq:self_sv1}
\Sigma^M_{s,v} &=&
g^2_M\int \frac{d\bm q}{(2\pi)^3}
\frac{\mathscr{Q}_{s,v}}{4E_qE_M} \nonumber\\
&\times&\left[\frac{i\omega_n \mathscr{C}_3-2E_+ \mathscr{C}_1}
{E_+^2+\omega_n^2}-
\frac{i\omega_n \mathscr{C}_3+2E_- \mathscr{C}_2}
{E_-^2+\omega_n^2}\right],\\
\label{eq:self_01}
\Sigma^M_0 &=&
g^2_M\int \frac{d\bm q}{(2\pi)^3} 
\frac{\mathscr{Q}_{0}}{4 E_qE_M}\nonumber\\
&\times&\left[\frac{2i\omega_n \mathscr{C}_1-E_+\mathscr{C}_3}{E_+^2+\omega_n^2}+
\frac{2i\omega_n \mathscr{C}_2+E_-\mathscr{C}_3}
{E_-^2+\omega_n^2}\right],
\eea
with the short-hand notations
\bea
&&\mathscr{C}_1=1+n_B(E_M)-\frac{1}{2}[n^+(E_q)+n^-(E_q)],\nonumber\\
\label{eq:Z_123}
&&\mathscr{C}_2=n_B(E_M)+\frac{1}{2}[n^+(E_q)+n^-(E_q)],\\
&&\mathscr{C}_3=n^+(E_q)-n^-(E_q),\nonumber
\eea
and (see Appendix \ref{app:C} for details)
\bea\label{eq:f_sv}
\mathscr{Q}_{s}=1,\quad
\mathscr{Q}_{v}=\frac{\bm q\cdot\bm p}{p^2},\quad
\mathscr{Q}_{0}=-\frac{E_q}{i\omega_n}.
\eea
The retarded self-energy is now obtained by analytical continuation
$i\omega_n\to p_0+i\varepsilon$ and has the same Lorentz structure as
its Matsubara counterpart.  To obtain the quark self-energy one has to
take into account the contributions from three pions and one $\sigma$-meson. 
The components of this self-energy sum up as follows
\bea\label{eq:selfsum2} 
\Sigma_s= \Sigma^\sigma_s-3\Sigma^\pi_s,\quad
\Sigma_{0/v} =-\Sigma^\sigma_{0/v} -3\Sigma^\pi_{0/v}.
\eea
For the imaginary part of the on-shell quark 
self-energy
($\varrho \equiv {\rm Im}\Sigma$) one finds
\bea\label{eq:im_self}
\!\!\varrho_{j}^M(p)\Big\vert_{p_0=E_p}\!\!\! =
\frac{g^2_M}{16\pi p}
\int_{E_{\rm min}}^{E_{\rm max}}\!\!\! d E
\mathscr{T}_{j}[n_B(E_M)+n^-(E)],
\eea
where $j=s,0,v,$  $E_M=E+E_p$, and
\bea\label{eq:f_sv}
\mathscr{T}_{s}=1,\quad
\mathscr{T}_{v}=\frac{m_M^2-2m^2-2EE_p}{2p^2},\quad
\mathscr{T}_{0}=-\frac{E}{E_p}.
\eea
The distribution function 
for quarks and antiquarks is defined as 
$n^\pm(E)=[e^{\beta(E \mp \mu)}+1]^{-1}$, and 
$n_B(E)=(e^{\beta E}-1)^{-1}$ is the Bose distribution
function for zero chemical potential.
In the same way we find for the antiquark on-shell 
self-energy ($p_0=-E_p$) 
\bea\label{eq:im_self_anti}
\varrho^M_{j}(p)\Big\vert_{p_0=-E_p}\!\!\!\!=-
\frac{g^2_M}{16\pi p}
\int_{E_{\rm min}}^{E_{\rm max}}\!\!\!\!d E
\mathscr{T}_{j}[n_B(E_M)+n^+(E)],\nonumber\\
\eea
 where
\bea\label{eq:E_min_max}
 E_{{\rm min},{\rm max}}&=&
\frac{1}{2m^2}\left[(m_M^2-2m^2)  p_0 \right. \nonumber \\
&  & \hspace*{1cm} \left. \pm pm_M\sqrt{m_M^2-4m^2}\right].
\eea  
 The range of integration is according to Eq.~\eqref{eq:E_min_max} 
\bea\label{eq:E_range}
 E_{\rm max}-E_{\rm min}=\frac{pm_M}{m^2}\sqrt{m_M^2-4m^2}. 
\eea 
We note that if $m_M\ge 2m$ the
integration range is not empty.  If $m=0$, we have
\bea\label{eq:E_min_max_chiral}
 E_{\rm min}=\frac{m_M^2}{4p},\quad
 E_{\rm max}\to \infty.
\eea
The full quark-antiquark self-energy in on-shell approximation can be written as
\bea\label{eq:im_self_onshell}
\varrho_j(p_0,p) =
\theta (p_0)\varrho^+_j(p)+
\theta (-p_0)\varrho^-_j(p),
\eea
with $\varrho^\pm_j(p)=\varrho_j(p_0=\pm E_p,p)$.
It is seen from Eqs.~\eqref{eq:im_self} and \eqref{eq:im_self_anti} that
$\varrho^+$ and $\varrho^-$ are related by the relation
$\varrho^-_j(\mu, p)=-\varrho^+_j(-\mu, p)$, therefore
\bea\label{eq:relation_rho}
\varrho_j(\mu, -p_0, p)&=&\theta (-p_0)\varrho^+_j(\mu, p)+
\theta (p_0)\varrho^-_j(\mu, p)\nonumber\\
&=&-\theta (-p_0)\varrho^-_j(-\mu, p)-
\theta (p_0)\varrho^+_j(-\mu, p) \nonumber\\
&=&-\varrho_j(-\mu, p_0, p),
\eea
where we indicated the $\mu$-dependence 
of the self-energy explicitly.
\begin{figure*}[!] 
\begin{center}
\includegraphics[width=14.0cm,keepaspectratio]{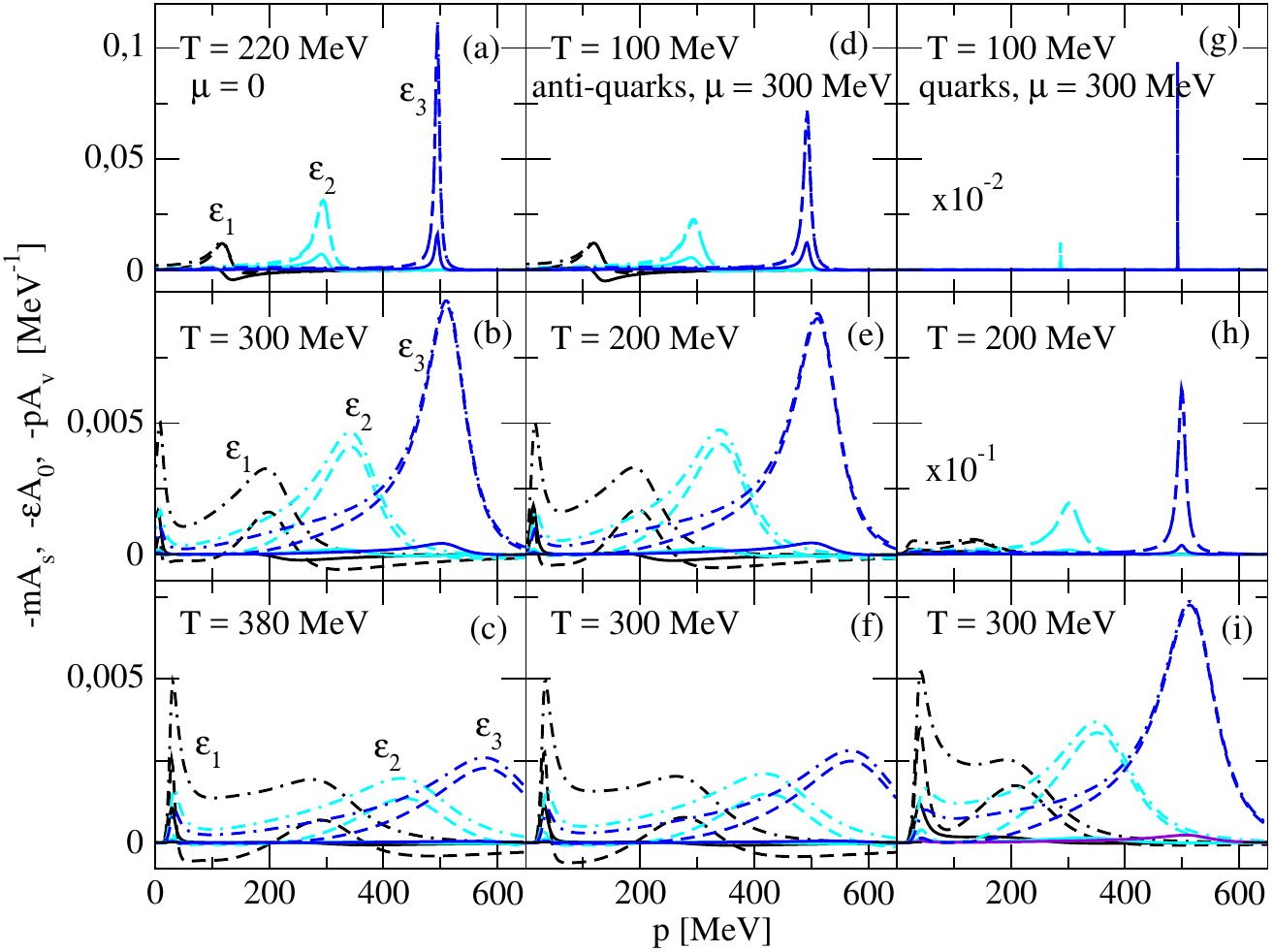}
\caption{ Dependence of three Lorentz components of the quark and
  antiquark spectral functions $-mA_s$ (solid line),
  $-\varepsilon A_0$ (dash-dotted line) and $-pA_v$ (dashed line) on
  the momentum.  Figures (a)-(c) correspond to $\mu=0$, (d)-(f) to
  antiquarks with $\mu=300$ MeV, and (g)-(i) to quarks with $\mu=300$
  MeV. These spectral functions are shown at three energies
  $\varepsilon_1 = 100$, $\varepsilon_2 = 300$, and
  $\varepsilon_3 = 500$ MeV, as indicated in the plot. Note that the
  vector component of the spectral function changes its sign, whereas
  the remaining components do not, see the discussion in the text.  }
\label{fig:spectral} 
\end{center}
\end{figure*}

Now for any transport coefficient $\chi(\mu)$
 we can write
\bea\label{eq:coeff}
\chi(\mu)=\int_{-\infty}^{\infty} 
d\varepsilon n^+(\varepsilon)[1-n^+(\varepsilon)]
\int_{0}^{\Lambda} dp
\mathscr{F}(p,\varepsilon,\mu),
\eea
where
$\mathscr{F}(p,\varepsilon,\mu)\equiv
\mathscr{F}(p,\varepsilon,h(\mu),\varrho_{j}(\mu,\varepsilon,p))$
is an even function of $p$ and $\varrho_{j}$, as seen from
Eqs.~\eqref{eq:kappafinal}, \eqref{eq:sigmafinal}, \eqref{eq:eta2}, and
\eqref{eq:spectral_coeff}-\eqref{eq:N2}.  It is invariant under the
inversion $\varepsilon\to -\varepsilon$ in the cases of the electrical
conductivity and the shear viscosity, and under the simultaneous
inversions $\varepsilon\to -\varepsilon$, $h\to -h$ in the case of the
thermal conductivity.  Because $h$ is an odd function of the chemical
potential, see Appendix \ref{app:D}, 
it follows from Eq.~\eqref{eq:relation_rho}  that  ($\varepsilon \equiv p_0$)
\bea
&&\mathscr{F}(p,-\varepsilon,\mu)=\mathscr{F}(p,
-\varepsilon,h(\mu),\varrho_{j}(\mu,-\varepsilon,p)) 
\nonumber\\ 
&&\hspace{0.5cm}
=\mathscr{F}(p,\varepsilon, -h(\mu),
-\varrho_{j}(-\mu,\varepsilon,p)) 
\nonumber\\ 
&&\hspace{0.5cm}
=\mathscr{F}(p,
\varepsilon,h(-\mu),\varrho_{j}(-\mu,\varepsilon,p)) 
=\mathscr{F}(p,\varepsilon,-\mu).\nonumber
\eea
Using this property in combination with relations
$n^+(\mu,-\varepsilon)=1-n^-(\mu,\varepsilon)$,
$n^-(\mu,\varepsilon)=n^+(-\mu,\varepsilon)$, and employing
Eq.~\eqref{eq:im_self_onshell} we rewrite Eq.~\eqref{eq:coeff} as
\bea\label{eq:coeff1} \chi(\mu)=\chi^+(\mu)+\chi^-(\mu), \eea
where we separated the contributions from 
positive and negative energies
\bea\label{eq:coeff_plus} 
\chi^+(\mu) &=&\int_{0}^{\infty}
d\varepsilon n^+(\mu,\varepsilon)[1-n^+(\mu,\varepsilon)]\nonumber\\
&& \times \int_{0}^{\Lambda} dp
\mathscr{F}(p,\varepsilon,\varrho_{j}^+(\mu,p)),\\
\label{eq:coeff_minus}
\chi^-(\mu)&=&\int_{-\infty}^{0} d\varepsilon
n^+(\mu,\varepsilon)[1-n^+(\mu,\varepsilon)]
\nonumber\\ && \times \int_{0}^{\Lambda} dp
\mathscr{F}(p,\varepsilon,\varrho_{j}^-(\mu,p)) 
=\chi^+(-\mu).
\eea
Therefore from Eq.~\eqref{eq:coeff1} we obtain 
\bea\label{eq:coeff2}
\chi(\mu)=\chi^+(\mu)+\chi^+(-\mu),  
\eea 
which implies that the transport coefficients 
are even functions of
the chemical potential, as expected.

\section{Numerical results}
\label{sec:results}

\subsection{Self-energies and spectral functions}

The imaginary parts of the quark and antiquark on-shell
self-energies, given by Eqs.~\eqref{eq:im_self} and
\eqref{eq:im_self_anti}, respectively, are shown in
Fig.~\ref{fig:self_energy} as functions of the quark momentum $p$ at
fixed values of the temperature and the chemical potential. For each
value of $\mu$ the temperature values are chosen to cover the range
$T_{\rm M}< T< T_{\rm max}$, as displayed in
Fig.~\ref{fig:mott_temp1}. Below $T_{\rm M}$, which is defined by the
continuum condition $m_\pi = 2m$, the imaginary parts of the on-shell
self-energies of quarks are negligible, since the processes of quark
scattering with meson exchange are kinematically forbidden in the case
of $\pi$-mesons, and are strongly suppressed in the case of $\sigma$-mesons 
if compared to off-shell processes.

The three components of the self-energies differ by the factors
$\mathscr{T}_{j}$. As in the case of the scalar self-energy
$\mathscr{T}_{s}=1$, we conclude that the differences seen between the
components $\varrho_j$ in Fig.~\ref{fig:self_energy}, as for example
the sign change between $\varrho_s$ and $\varrho_0$ and more
pronounced maxima in $\varrho_0$ and $\varrho_v$ than in $\varrho_s$,
originate from the $\mathscr{T}_{j}$ factors in Eq.~\eqref{eq:f_sv}. To
understand the small-$p$ behavior of the self-energies, note first
that the explicit $p^{-1}$ divergence in Eqs.~\eqref{eq:im_self} and
\eqref{eq:im_self_anti} is cancelled by the linear-in-$p$ dependence
of the integration range, given by Eq.~\eqref{eq:E_range}.
Furthermore, when $p\to 0$ the limits of integration tend to
$(m_M^2-2m^2)/2m$, see Eq.~\eqref{eq:E_min_max}.
If this limiting value is comparable to the
temperature, then the quark and antiquark distribution functions are
nonzero and contribute to the self-energies for $p\to 0$; this is the
case in Figs.~\ref{fig:self_energy} (a), (d), and (g).  In case where
$(m_M^2-2m^2)/2m \gg T$ the small-$p$ contributions are suppressed by
the vanishingly small distribution functions, see Figs.~\ref{fig:self_energy} (b), (e), and (h), as well as
(c), (f), and (i).  However, in the chiral
limit $m_0=0$ we always find the asymptotic behavior
${\rm Im}\Sigma \to 0$, when $p\to 0$, because the lower bound of the
integral \eqref{eq:im_self} becomes infinitely large as seen from
Eq.~\eqref{eq:E_min_max_chiral}.

In the large-$p$ limit the integration range is broad and the
asymptotics is controlled by the cutoff of high-momentum contributions
by the distribution functions, as well as the
factors $\mathscr{T}_{j}$. Thus the appearance of the maxima in the
self-energies [necessarily in the case $(m_M^2-2m^2)/2m \gg T$] is the
consequence of this asymptotic behavior. The shifts of the maxima
to higher momenta with increasing temperature is caused by the shift
in energy sampling of the distribution functions.

Next we examine the self-energies at fixed chemical potential, \ie,
the vertical columns in Fig.~\ref{fig:self_energy}.  Because the difference
$m_M-2m$ increases with temperature as we move away from the Mott line
the integration region increases. At the same time the distribution
functions cover phase space with higher energies. In combination this leads
to an increase of the imaginary parts of self-energies of quarks and
antiquarks with temperature, which is well pronounced for high
momenta. This increase is also caused by the additional temperature
dependence of the coupling constant $g_M$, see
Fig.~\ref{fig:couplings} in Appendix \ref{app:B}.

Consider now the dependence of the self-energies on the chemical
potential for fixed temperature by comparing, for example, Figs.~\ref{fig:self_energy} (a), (e), and (h)
with (b), (f), and (i). We observe two effects:
(i) the contributions to the self-energies from large $p$ becomes
larger for $\mu \neq 0$ both for quarks and antiquarks; (ii) the
overall magnitude of the quark self-energies (for example the maximum) is
reduced for $\mu \neq 0$. Nonzero $\mu$ affects quark
self-energies stronger than that of antiquarks because of a
stronger depletion of the antiquark population at nonzero baryon
density, on which the quark self-energy depends.  In fact, at any
temperature the quark self-energies are by a factor of two smaller than
their antiquark counterparts for $\mu = 200$ MeV and this difference
grows for large $\mu$; e.g., for $\mu=300$ MeV the suppression factor
is ten at $T=100$ MeV.  A similar comparison for antiquarks [see
(a)-(c) and (d)-(f)] shows that the antiquark self-energies at
nonzero $\mu$ are comparable to the self-energies for the $\mu =0$ case.

\begin{figure}[!] 
\begin{center}
\includegraphics[width=8.0cm,keepaspectratio]{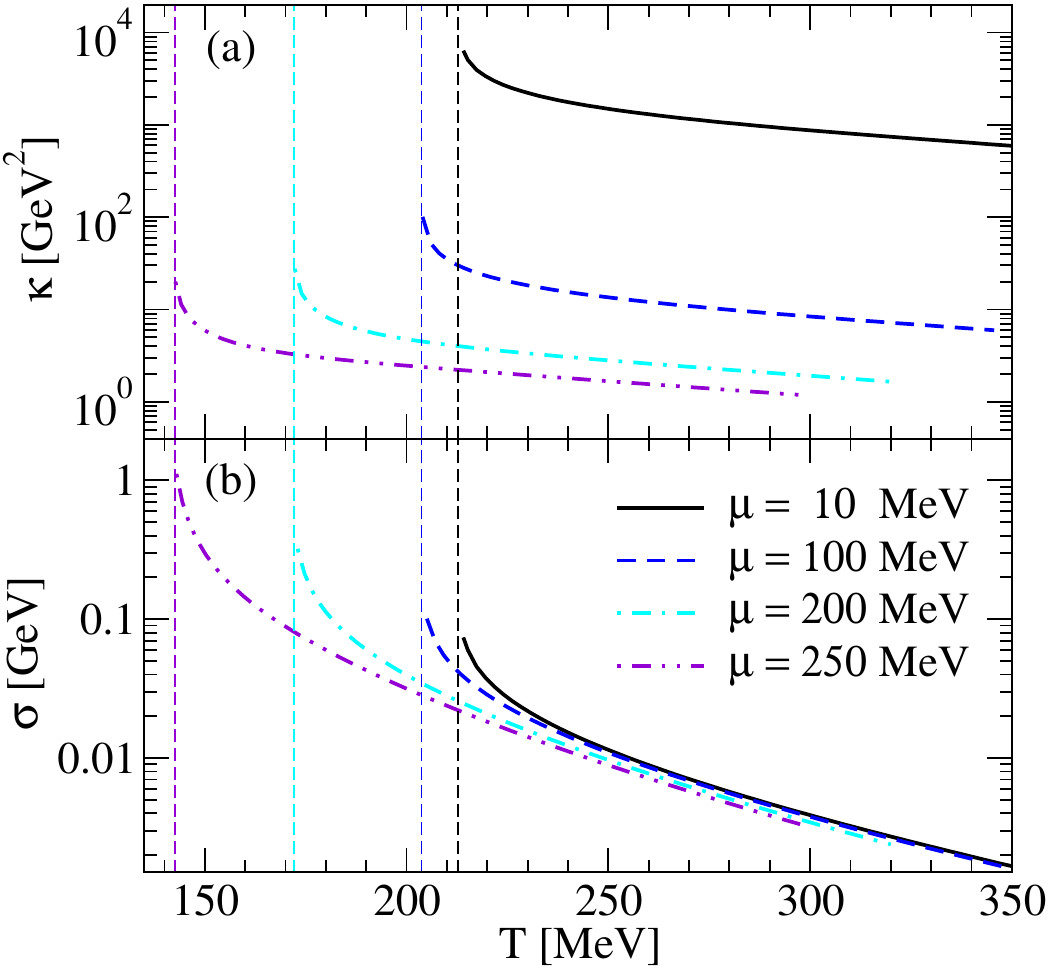}
\caption{ The temperature dependence of (a) thermal and (b) electrical
  conductivity at various values of the chemical potential. The
  vertical lines show the Mott temperature at the given value of
  $\mu$. }
\label{fig:kappa_sigma1} 
\end{center}
\end{figure}
Next we turn to the three Lorentz components of the spectral function
given by Eqs.~\eqref{eq:spectralfunction}-\eqref{eq:N2}, which are
shown in Fig.~\ref{fig:spectral} as functions of the quark momentum at
three values of the quark (off-shell) energy, $\varepsilon_1 = 100$,
$\varepsilon_2 = 300$, and $\varepsilon_3 = 500$ MeV.  The
quasiparticle peak in the spectral functions appears for
$p\simeq \varepsilon$, as expected from Eqs.~\eqref{eq:N1} and \eqref{eq:N2}.  An
estimate gives $n_1\approx (p_0^2-p^2)(1-{\varrho}_{0,v}^2)$ and
$n_2\approx (p_0^2+p^2){\varrho}_0$, therefore the denominator $d$
attains its minimum roughly at $p\simeq p_0$. In all cases it is seen
that the heights of the peaks increase with the quark energy. As
expected on physical grounds, the quasiparticle peaks are broadened
with increasing temperature and are replaced by more complex
structures in the high-temperature regime, see Figs.~\ref{fig:spectral} (c),  (f), and (i).

A comparison of quark and antiquark spectral functions shows that the
quasiparticle peaks of quarks are sharper than those of antiquarks for
the same temperature and chemical potential. As indicated in
Fig.~\ref{fig:spectral}, for $\mu=300$ MeV the peak in the spectral
function of quarks is by a factor of $10^2$ larger than that of antiquarks
at $T=100$ MeV and by factor of ten at $T=200$ MeV.  Finally note that the
temporal and vector components of the spectral function are of the
same order of magnitude and are almost coinciding at high energies,
whereas the scalar component is always suppressed. Thus, we may
already conclude that the main contribution to the transport
coefficients will originate from the temporal and vector components of
the spectral functions.  Note that the vector component of the
imaginary self-energy energy changes the sign, consequently the
corresponding spectral function changes its sign as well. However, 
the overall spectral width of the quasiparticles, which defines their
decay rate and contains contributions from all Lorentz
components remains positive~\cite{LangDiss}.
\begin{figure}[!] 
\begin{center}
\includegraphics[width=8.0cm,keepaspectratio]{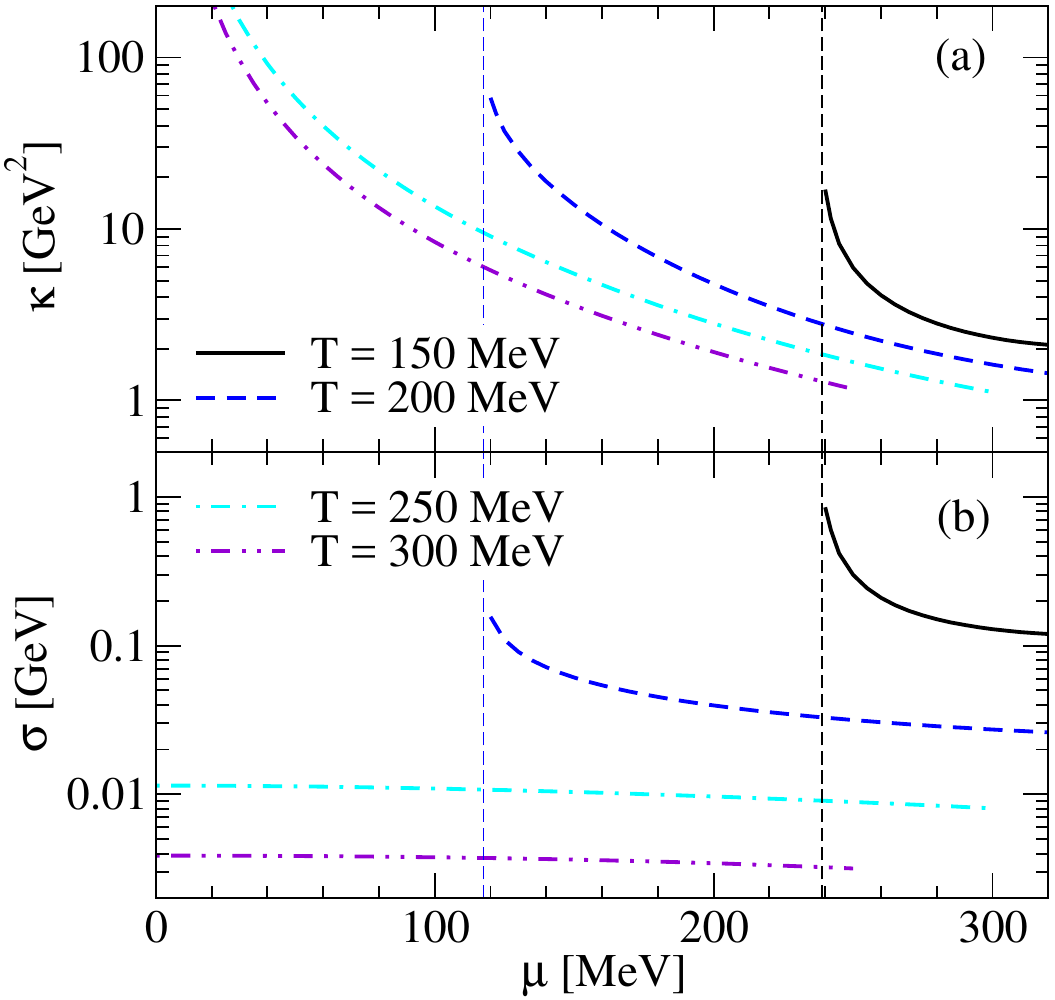}
\caption{ The dependence of (a) thermal  and (b) electrical conductivity
 on the chemical potential at various temperatures.
  The vertical lines show the value of the chemical potential where
  the temperature approaches the Mott temperature.  }
\label{fig:kappa_sigma2} 
\end{center}
\end{figure}
\begin{figure}[!] 
\begin{center}
\includegraphics[width=8.0cm,keepaspectratio]{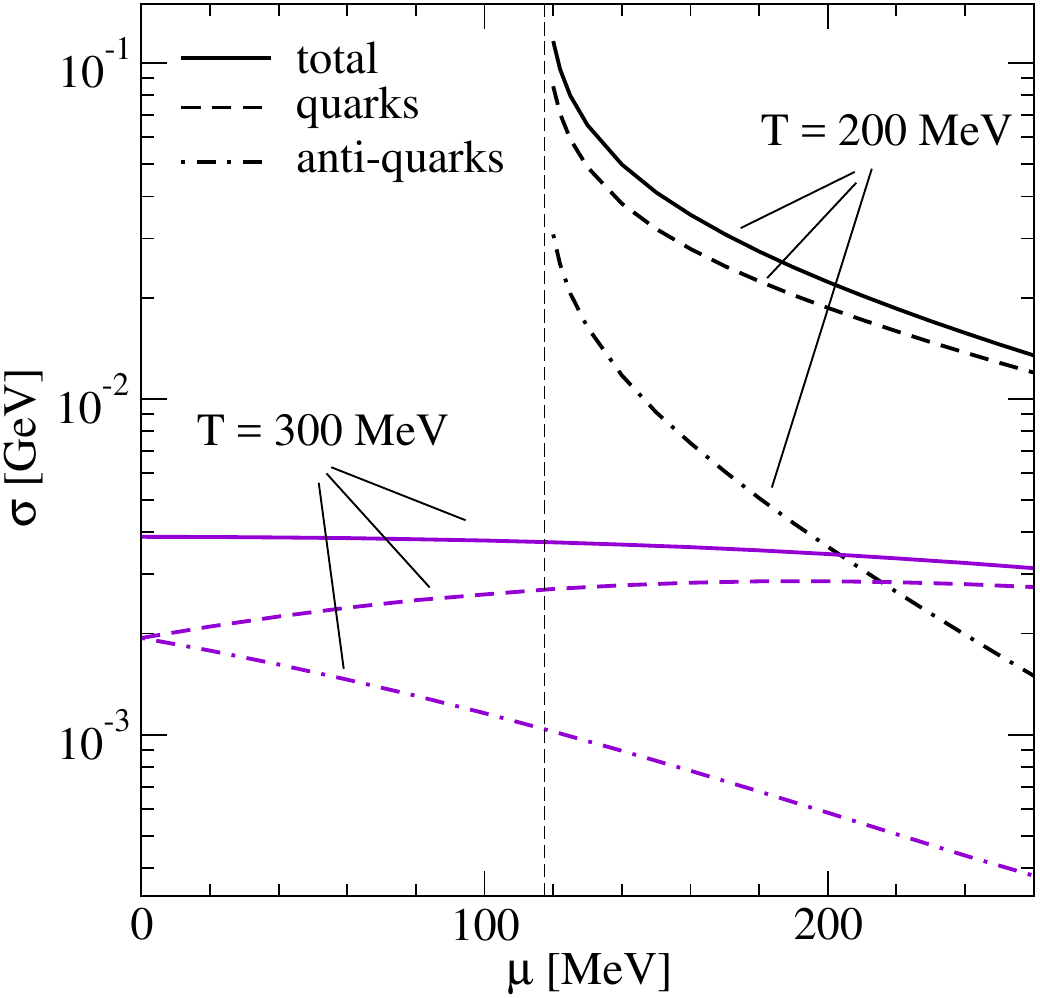}
\caption{ The partial contributions of quarks and antiquarks to the
  electrical conductivity and their sum as functions of the chemical
  potential at two temperatures indicated in the plot. }
\label{fig:sigma_partial} 
\end{center}
\end{figure}

\subsection{Thermal and electrical conductivities }

We now turn to the evaluation of the thermal and electrical
conductivities as given by Eqs.~\eqref{eq:kappafinal} and
\eqref{eq:sigmafinal}. Before discussing the numerical results,
consider the generic structure of these expressions.  For a given
value of $\varepsilon$ the inner integrand has a peak structure with a
maximum located at $p\simeq \varepsilon$, as implied by the shape of
the spectral functions.  The heights of the peaks rapidly increase
with $\varepsilon$.  As a consequence, the inner momentum integral in
Eqs.~\eqref{eq:kappafinal} and \eqref{eq:sigmafinal} is a rapidly
increasing function of $|\varepsilon|$ as long as
$\vert\varepsilon\vert\leq\Lambda$.  For energies larger than
$\Lambda$ the peaks are outside of the integration range (because of
the momentum cutoff) and the integral sharply decreases with
$\varepsilon$. The outer integration contains the factor
$\partial n(\varepsilon)/\partial\varepsilon$ which at low
temperatures is strongly peaked at the energy $\varepsilon = \mu$. At
high temperature it transforms into a broad, bell-shaped structure
which samples energies far away from $\mu$.

It is evident from Eqs.~\eqref{eq:kappafinal} and
\eqref{eq:sigmafinal} that for $\mu \to 0$ the integrands of the momentum
integrals are even functions of $\varepsilon$, as discussed above, and
the quark and antiquark contributions originating from positive and
negative ranges of the $\varepsilon$-integration are equal. At nonzero
chemical potentials the contribution of antiquarks is suppressed by
both spectral functions and by the factor
$\partial n/\partial\varepsilon$. We will give an explicit numerical
example below in Fig.~\ref{fig:sigma_partial}.

Figure \ref{fig:kappa_sigma1} shows the temperature dependence of
$\kappa$ and $\sigma$ for several values of the chemical
potential. The conductivities
decrease with temperature for all values of the chemical potential. The observed decrease is the result of the
broadening of the spectral functions with temperature, which
physically corresponds to stronger dispersive effects and shorter
relaxation times. This implies smaller conductivities.

Note that at the Mott temperature and below the conductivities become
very large because the dispersive effects incorporated in the spectral
functions via the imaginary parts of the self-energies vanish for
pions and are very small for the $\sigma$-meson. This is the
consequence of the on-shell approximation, and can be improved if one
incorporates off-shell contributions to the self-energies. This
improvement close to (and below) the Mott temperature that
incorporates off-shell kinematics is unimportant at temperatures
already slightly above the Mott temperature, where the transport
coefficients are described by on-shell kinematics quite well. (For a
computation of off-shell self-energies see Ref.~\cite{LKW15}, where their
impact for the shear viscosity were found to be small.)
\begin{figure}[!] 
\begin{center}
\includegraphics[width=8.0cm,keepaspectratio]{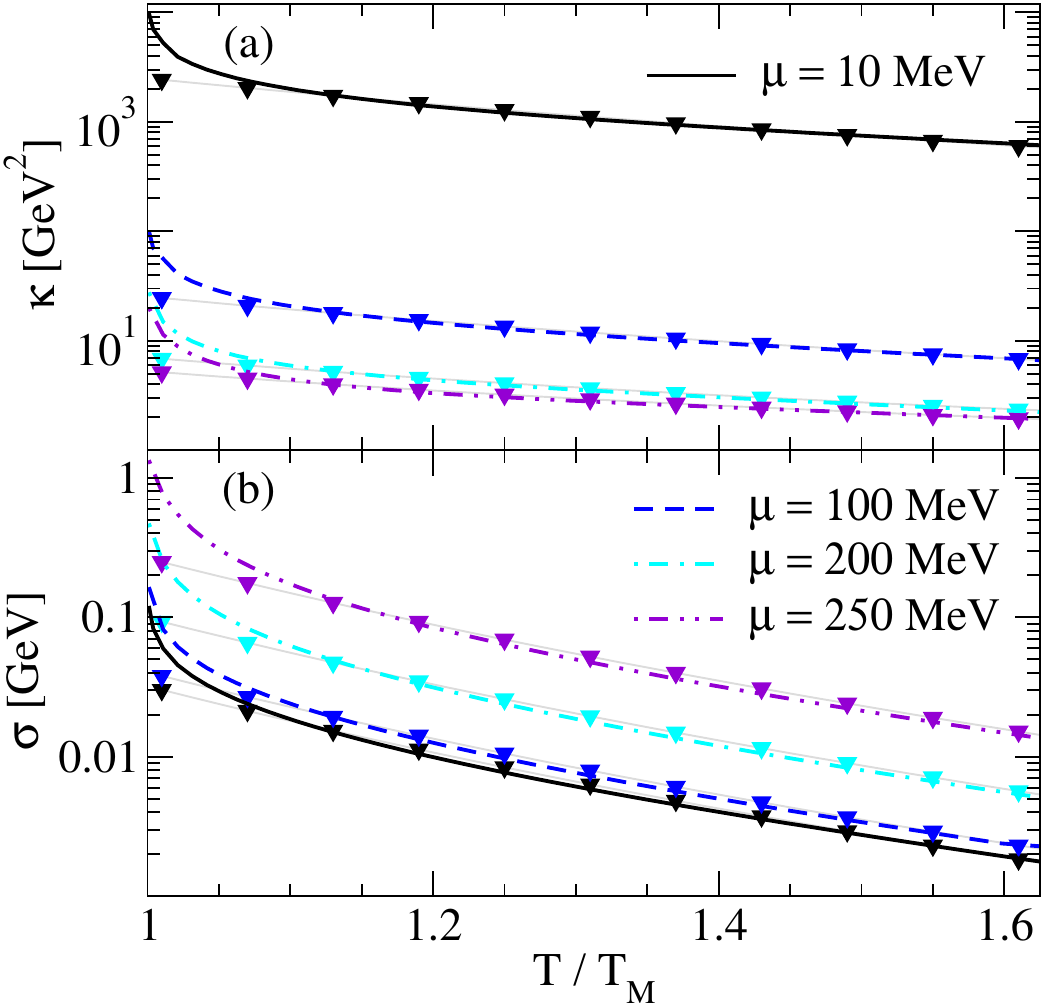}
\caption{ The thermal conductivity $\kappa$ (a) and electrical
  conductivity $\sigma$ (b) as functions of the scaled temperature
  $T/T_{\rm M}$ at several values of the chemical potential.}
\label{fig:kappa_sigma3} 
\end{center}
\end{figure}
\begin{figure}[!] 
\begin{center}
\includegraphics[width=8.2cm,keepaspectratio]{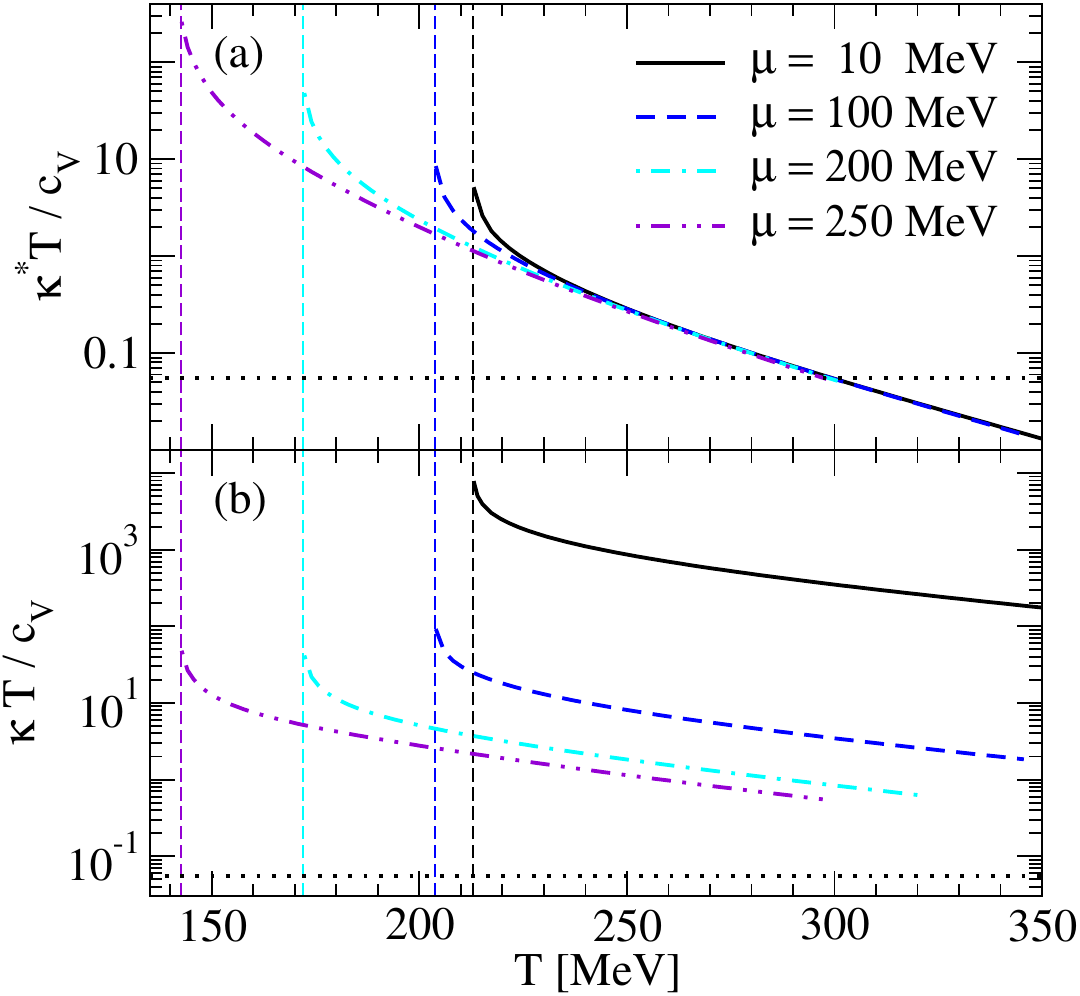}
\caption{ The ratios $\kappa^* T/c_V$ (a) and $\kappa T/c_V$ (b) as
  functions of the temperature at several values of the chemical
  potential.}
\label{fig:kappa_sigma4} 
\end{center}
\end{figure}

Comparing the overall behavior of the thermal and electrical
conductivities we observe two main differences: (i) the electrical
conductivity drops faster with temperature than the thermal
conductivity, (ii) for small chemical potentials the thermal
conductivity diverges, whereas the electrical conductivity remains
almost independent of the chemical potential.  Both effects originate
from those terms in Eq.\ \eqref{eq:kappafinal} for $\kappa$ which
contain the enthalpy $h$.  In the relevant temperature-density range,
the minimal value of the enthalpy per particle is
$h_{\rm min}\simeq 0.8$ GeV, see Fig.~\ref{fig:enthalpy} in
Appendix \ref{app:D}. This value already exceeds the cutoff parameter
$\Lambda\simeq 0.65$ GeV, which is the characteristic energy scale of
the model, therefore one may conclude that the dominant terms in
$\kappa$ are the terms containing $h$, \ie, the terms arising from
the second and third correlators on the right-hand side of
Eq.~\eqref{eq:corkappa_m}. The enthalpy per particle rapidly increases
with the decrease of the chemical potential, therefore at small
chemical potentials the main contribution comes from the third
term. Numerically we find that the first two terms are negligible
compared to the third one for $\mu\le 100$ MeV.  The second correlator
becomes important once $\mu\ge 100$ MeV, whereas the first one is
always suppressed for $\mu\le 250$ MeV. Thus, using
Eqs.~\eqref{eq:kappa}, \eqref{eq:sigma}, \eqref{eq:corkappa_m}, and
\eqref{eq:feynmanrules4}, we obtain for small chemical potentials
$\mu\le 100$ MeV a simple relation between thermal and electrical
conductivities
\bea\label{eq:kappa_sigma_ratio}
\frac{\kappa}{\sigma}=\frac{N_f}{\Tr\hat Q^2}
\frac{h^2}{T}=\frac{9h^2}{10\pi\alpha T}.
\eea 
Note that the first equality holds for any number of flavors, whereas
in the second step we substituted $N_f = 2.$ For $\mu\ll T$ we have the
asymptotic behavior $h\simeq 7\pi^2 T^2/15\mu$, therefore the thermal
conductivity diverges at vanishing baryon density as
$\kappa\propto\mu^{-2}$~\cite{1985PhRvD..31...53D}.  Substituting the
expression for $h$ in Eq.~\eqref{eq:kappa_sigma_ratio} we find in the
nondegenerate regime 
\bea\label{eq:kappa_sigma_ratio1} \frac{\kappa
  \mu^2}{\sigma T^3}= \frac{N_f}{\Tr\hat
  Q^2}\left(\frac{7\pi^2}{15}\right)^2=\frac{49\pi^3}{250\alpha}\simeq
830.  
\eea 
It is seen that the Wiedemann-Franz law $\sigma T/\kappa \sim$ const.
does not hold in this case.  Finally, we note that away from the Mott line
we have the scalings $\sigma\propto T^{-6}$ and
$\kappa\propto T^{-\gamma}$ with $\gamma =3$ for $\mu \le 100$ MeV and
$\gamma =2$ for $\mu \ge 200$ MeV.

The dependence of conductivities on the chemical potential is shown in
Fig.~\ref{fig:kappa_sigma2}. The electrical conductivity is seen to be
nearly independent of the chemical potential away from the Mott
transition line. Only close to this transition $\sigma$ increases
because of the vanishing of the spectral width at $T_{\rm M}$. Note
that for temperatures $T\ge 250$ MeV the electrical conductivity
remains almost constant because of the absence of the Mott line at
these temperatures.  However, the thermal conductivity is always a
rapidly decreasing function of the chemical potential because of the
reasons discussed above, and becomes infinitely large in the limit
$\mu\to 0$.

At nonzero $\mu$ the symmetry between quarks and antiquarks is
broken; its consequences discussed above are illustrated in
Fig.~\ref{fig:sigma_partial}, where the quark and antiquark
contributions to $\sigma$ are shown separately. At temperatures close
to $T_{\rm M}$ both contributions decrease with quarks contributing
dominantly. For temperatures away from the Mott transition line, the moderate increase in
the conductivity of quarks up to $\mu \simeq 200$ MeV is accompanied
by a rapid decrease in the contribution of antiquarks, which
becomes negligible at $\mu > 200$ MeV. The sum of these two
contributions turns out to be a slowly decreasing function of $\mu$ in
the entire range of $\mu$ and $T$, see Fig.~\ref{fig:kappa_sigma2}.

In order to remove the effect of the variations of the position of the
Mott line with the chemical potential, we show in
Fig.~\ref{fig:kappa_sigma3} again the conductivities as in
Fig.~\ref{fig:kappa_sigma1}, but with the temperature axis scaled by the
corresponding $T_{\rm M}(\mu)$. In this case the conductivities display a
universal dependence on this scaled temperature, their values being
only shifted by a $\mu$-dependent constant.

Apart from the Wiedemann-Franz relation, the ratios the $\kappa T/c_V$
and $\kappa^{*} T/c_V$ are of interest. Here $\kappa^{*}$ is defined
by Eq.~\eqref{eq:kappafinal} with $h=0$ and $c_V$ is the specific heat
capacity as defined in Appendix \ref{app:D}.  These ratios are shown
in Fig.~\ref{fig:kappa_sigma4}.

We conjecture that the ratio $\kappa^{*} T/c_V$, which is associated
with the energy transfer, is bounded from below due to the quantum
mechanical uncertainty principle.  To motivate this conjecture we
refer to kinetic theory of noninteracting gases.  Because for
temperatures $T> T_{\rm M}$ the quark masses are negligible compared
to other scales, we can set their average velocity $\bar{v}\simeq
1$.
Furthermore, in this regime the characteristic energy
$\varepsilon\simeq 3T$. Now, according to the kinetic theory of dilute
gases the thermal conductivity is estimated as
\bea\label{eq:ratio1}
\kappa^{\rm kin}\simeq\frac{1}{3} c_V\bar{v}l \simeq
\frac{1}{3}c_V\tau,
\eea
where $l$ is the mean free path and $\tau=l/\bar{v}$ is the mean
collision time. Therefore, we find that at high temperatures
$\kappa T/c_V \simeq T\tau/3 \simeq \varepsilon \tau/9$. Because of
the uncertainty principle $\varepsilon \tau \ge 1/2$, we find the
following bound (recovering the natural constants)
\bea\label{eq:bound}
\frac{\kappa^{\rm kin} T}{c_V} \ge \frac{\hbar c^2}{18k_B}.
\eea
In the present context $\kappa^{\rm kin}$ in Eq.\ \eqref{eq:bound}
should be associated with $\kappa^{*}$ because this quantity serves as
the analogue of the thermal conductivity defined in the kinetic theory
of gases. Indeed, both quantities involve the total energy transport
and do not separate the convective particle current which is excluded
from $\kappa$ by definition. 

Note that in the case of shear viscosity the discussion of an
analogous bound was given by Ref.~\cite{1985PhRvD..31...53D} and a
more stringent limit (the so-called KSS bound) was suggested later on
from gauge-gravity duality considerations~\cite{2005PhRvL..94k1601K}.

According to panel (a) of Fig.~\ref{fig:kappa_sigma4} the bound
\eqref{eq:bound} is violated for the ratio involving $\kappa^*$
at high temperatures $T> 300$ MeV. It is remarkable that this occurs
in the same range where the shear-viscosity bound is violated (see
below, Sec.~\ref{sec:shear}). The inclusion of gluonic degrees of
freedom will mitigate this violation. The quark-meson exchange
processes will not be the most dominant processes away from the Mott
temperature and the gluonic degrees of freedom are expected to play a
significant role in thermal transport.
\begin{figure}[!] 
\begin{center}
\includegraphics[width=8cm,keepaspectratio]{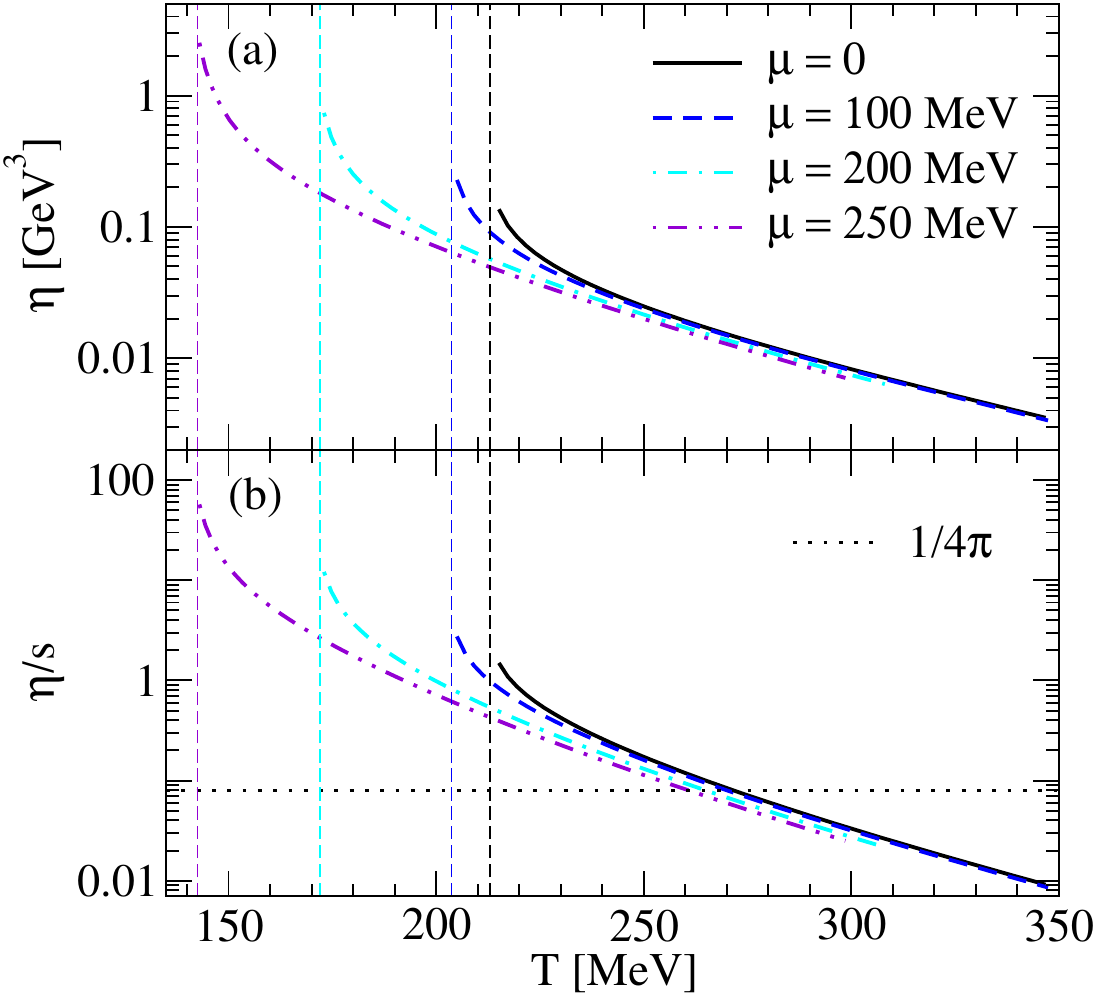}
\caption{ The temperature dependence of the shear viscosity $\eta$ (a)
  and its ratio to the entropy density (b) at various values of the
  chemical potential.  The vertical lines show the Mott temperature at
  the given value of $\mu$. }
\label{fig:shear1} 
\end{center}
\end{figure}
\begin{figure}[!] 
\begin{center}
\includegraphics[width=8.0cm,keepaspectratio]{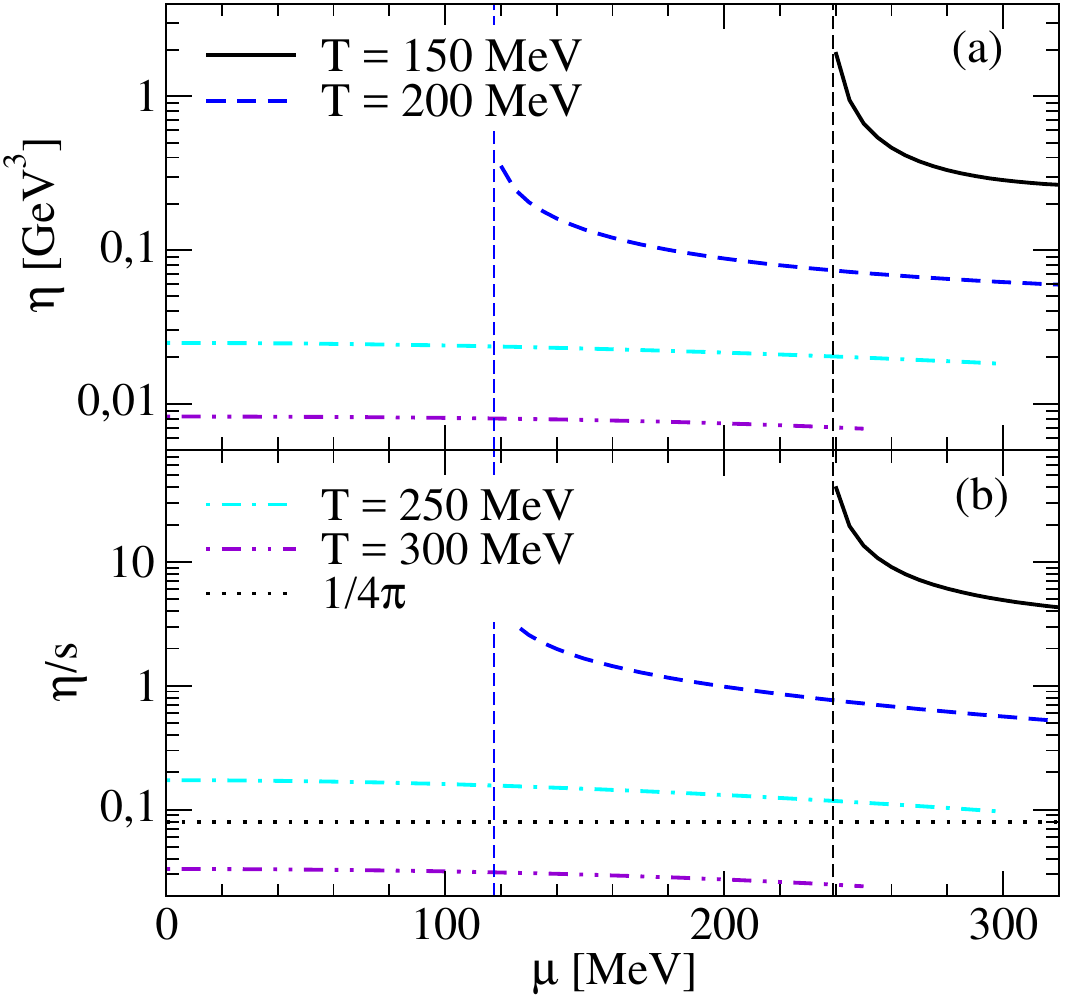}
\caption{ The dependence of the shear viscosity (a) and the ratio
  $\eta/s$ (b) on the chemical potential at various temperatures.  The
  vertical lines show the value of the chemical potential where the
  temperature approaches the Mott temperature. }
\label{fig:shear2} 
\end{center}
\end{figure}

\begin{figure}[!] 
\begin{center}
\includegraphics[width=8.0cm,keepaspectratio]{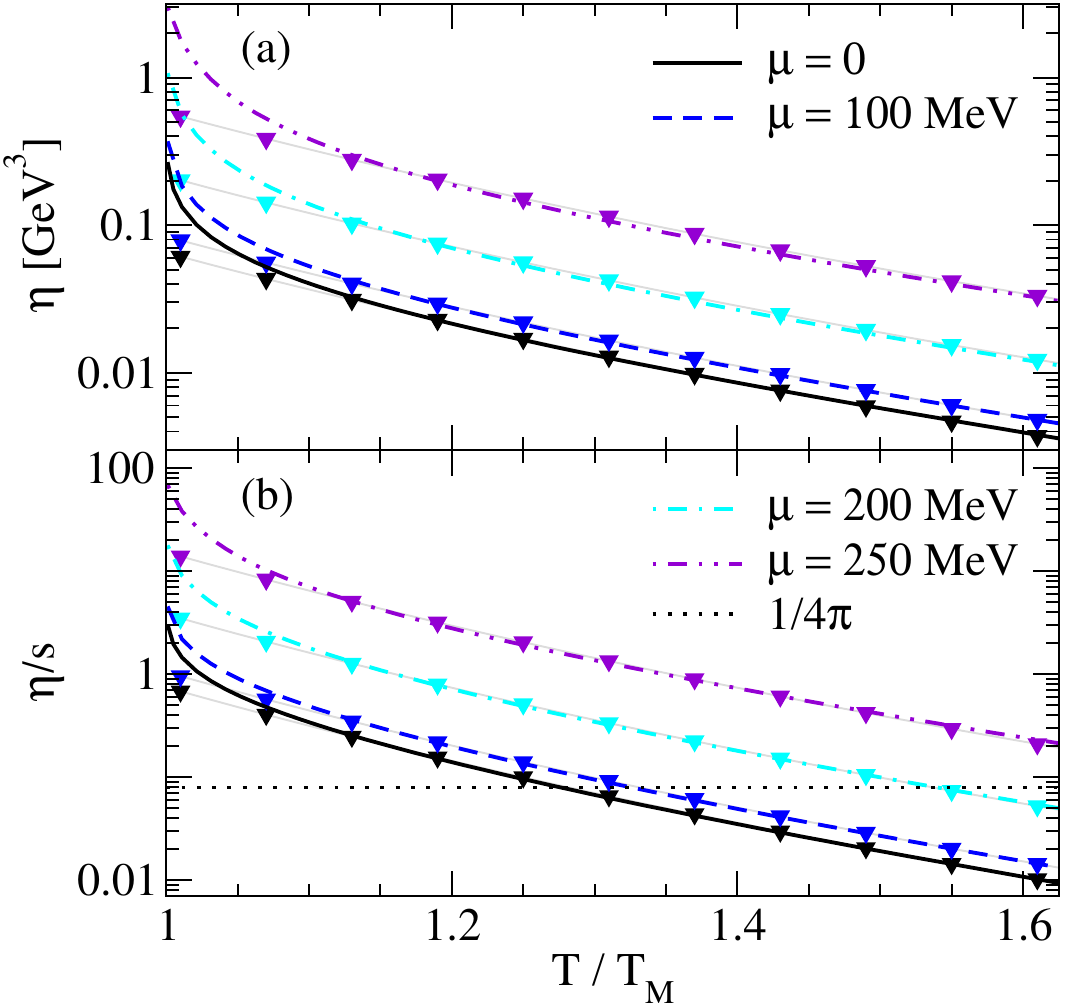}
\caption{The shear viscosity (a) and the ratio $\eta/s$ (b) as
  functions of the scaled temperature $T/T_{\rm M}$ at several values of the
  chemical potential. }
\label{fig:shear3} 
\end{center}
\end{figure}

\subsection{Shear viscosity}
\label{sec:shear}

The shear viscosity of quark matter has been studied extensively
because of the experimental evidence for its very low value in
heavy-ion collisions where quark-matter formation is expected and
because of the conjectured  universal lower bound of the ratio
$\eta/s$ derived from gauge-gravity duality.  We now
evaluate the expression for $\eta$ given by Eq.~\eqref{eq:eta2} and
compare it to earlier studies, in particular those based on the
two-flavor NJL model.

Figure \ref{fig:shear1} shows the temperature dependence of the shear
viscosity and the ratio $\eta/s$. The
entropy density of the present model is discussed in Appendix \ref{app:D}. As
in the case of the conductivities, the shear viscosity is a decreasing
function of the temperature, for the reasons already explained in
detail above: the dispersive effects increase with the temperature and
the viscosity of matter is reduced. The entropy density of quark matter is an
increasing function of temperature (linear in the degenerate regime
$T\ll \mu$ and cubic in the nondegenerate regime $T\gg \mu$),
therefore the ratio $\eta/s$ decreases faster than $\eta$ with increasing
temperature.  In the high-temperature regime $\eta$ and $\eta/s$ have
the scaling $T^{-6}$ and $T^{-9}$, respectively.  The enhancement of
both quantities as $T\to T_{\rm M}$ is understood as due to vanishing
of the relevant on-shell self-energies at the Mott temperature.  At
high temperatures $T\gtrsim 270$ MeV the ratio $\eta/s$ undershoots
the KSS bound 1/4$\pi$~\cite{2005PhRvL..94k1601K}. This is an
indication of the change in the processes that dominate the
viscosity, namely from quark-meson fluctuations to those including gluonic
degrees of freedom, which are integrated out from the NJL model.
Their contribution becomes increasingly important at high temperatures
and leads to an increase of the viscosity with temperature, see
Refs.~\cite{2000JHEP...11..001A,2008PhRvL.100q2301X,2015PhRvL.115k2002C,2003JHEP...05..051A}.

The dependence of $\eta$ and $\eta/s$ on the chemical potential is
shown in Fig.~\ref{fig:shear2}. As in the case of the electrical conductivity, these
are slowly decreasing functions of $\mu$ at fixed temperature except
at the corresponding Mott line where $\eta$ formally diverges.
Finally, in Fig.~\ref{fig:shear3} we show these quantities as functions
of the ratio $T/T_{\rm M}$. As in the case of the conductivities, we
observe a universal behavior of $\eta$ and $\eta/s$ on
$T/T_{\rm M}$ for fixed
$\mu$ values, \ie,  the curves belonging to different values of $\mu$ are
only shifted vertically by a $\mu$-dependent constant.
 
\subsection{Fitting transport coefficients}

The observed nearly universal behavior of the transport coefficients
with the scaled temperature $T/T_{\rm M}$ for fixed values of the chemical
potential suggests fitting transport coefficients as functions of $T/T_{\rm M}$ and the chemical potential, as displayed in
Figs.~\ref{fig:kappa_sigma3} and \ref{fig:shear3}.

For this purpose we first fit the Mott temperature, 
displayed in Fig.~\ref{fig:mott_temp1}, with the formula 
\begin{equation}\label{eq:fit_mott}
T_{\rm M}^{\rm fit}(\mu)=T_0
\left\{\begin{array}{ll} 1-\sqrt{\gamma y}
e^{-\pi/(\gamma y)}         &0\leq y\leq 0.5,\\
         \sqrt{1.55(1-y)+0.04(1-y)^2}  &0.5< y\leq
                                          1,\end{array}\right.
\end{equation}
with $T_0=T_{\rm M}(\mu=0)=213$ MeV, $y=\mu/\mu_0$, 
where $\mu_0=345$ MeV
corresponds to the point where $T_{\rm M}=0$ and the chemical potential
attains its maximum on the Mott line, and $\gamma =2.7$. 
The formula \eqref{eq:fit_mott} has relative accuracy $\le 3\%$ for  chemical potentials $\mu\le 320$ MeV.
 
Next, all 
transport coefficient can be fitted with a generic formula
\bea\label{eq:fit_formula}
\chi_{\rm fit} = C\left(\frac{T}{T_{\rm M}}\right)^{-\alpha}\exp[a_1y^2+a_2y^4+a_3y^6]\times
\chi_{\rm div},
\eea
where $\chi_{\rm fit} \in \{\sigma, \kappa, \kappa^*, \eta,
\eta/s\}$.
The term $\chi_{\rm div}$ is diverging in the limit $\mu\to 0$ in the
case of $\kappa$ and is given by the formula
\bea\label{eq:fit_kappa}
\chi_{\rm div} =
\left(\frac{T}{T_{\rm M}}\right)^{2}+y^{-2}.
\eea
For all other coefficients $\chi_{\rm div}=1$.  The values of the
constants in formula \eqref{eq:fit_formula} are given in 
Table~\ref{tab:1} (for each transport coefficient $C$ is given in
relevant units).
\begin{table}
\begin{tabular}{cccccc}
\hline
$\chi_{\rm fit}$       &$\quad C\quad$ & $\quad\alpha\quad$ &  $\quad a_1\quad$ &  
$\quad a_2\quad$ &  $\quad a_3\quad $\\
\hline 
$\sigma$& 0.032 & 6 &  2.64 &  1.23 &  2.67\\
$\kappa$& 2.10  & 3 &  -0.95 &  1.27 &  0.0\\
$\kappa^*$& 1.55  & 7 &  3.47 &  1.08 &  3.34\\
$\eta$    & 0.065  & 6 &  2.92 &  0.95 &  2.7\\
${\eta}/{s}$& 0.75 & 9 &  3.89 &  1.72 &  3.47\\
\hline
\end{tabular}
\caption{The values of the fit parameters in
  Eq.~\eqref{eq:fit_formula}.
}\label{tab:1}
\end{table}
A comparison between the exact results and the fits is shown
in Figs.~\ref{fig:kappa_sigma3} and \ref{fig:shear3}, where an excellent
agreement is observed for temperatures above $T/T_{\rm M} \ge 1.1$.
In this domain all fit formulas have relative accuracy $\le 10\%$.

\section{Conclusions}
\label{sec:conclusions}

In this work we have studied the electrical and thermal conductivities
as well as provided an update on the shear viscosity of quark matter
within the two-flavor NJL model using the Kubo-Zubarev
formalism~\cite{1957JPSJ...12..570K,zubarev1997statistical}. We have
derived Kubo formulas for the electrical and thermal conductivities of
a relativistic quark plasma taking into account the full Lorentz
structure of the self-energies (spectral functions) of the quarks. The
two-point correlation functions are evaluated with the full propagator
and within the $1/N_c$ approximation to the multiloop contributions;
these then imply that vertex corrections are suppressed and the
leading-order contributions to the correlation functions arise from
single-loop diagrams. It is worthwhile to note that our Kubo formulas
for the conductivities have generic validity and can be applied in the
broader context of field theories of relativistic plasmas, in a
straightforward manner when the vertex corrections are suppressed by
some mechanism. We have also revised the corresponding Kubo formula
for the shear viscosity.

We have applied this general formalism to compute the electrical and
thermal conductivities of the NJL model for quark matter in the regime
where the dispersive effects arise from quark-meson scattering above
the Mott temperature for dissolution of mesons into quarks. We find
that the conductivities are decreasing functions of temperature at
fixed chemical potential; they show nearly universal behavior when
temperature is scaled by the Mott temperature, \ie, as functions of
$T/T_{\rm M}$. We find that the ratio $\kappa /\sigma$ does not follow
the Wiedemann-Franz law. We then moved on to recompute the shear
viscosity of the model with our derived Kubo formula; we find a
qualitative agreement with previous results of
Refs.~\cite{2008JPhG...35c5003I,2008EPJA...38...97A,LW14,LKW15}.  In
particular the ratio of $\eta/s$ tends to the KSS bound
$1/4\pi$~\cite{2005PhRvL..94k1601K} but undershoots this bound at some
intermediate temperature and fails to describe the high-temperature
limit where gluonic degrees become important. We have also conjectured
a lower bound on the ratio $\kappa^* T/c_V> 1/18$. This conjecture
needs further studies from different standpoints, including those well
suited for nonperturbative calculations.  Within the NJL model we
find that this ratio undershoots the lower bound $1/18$ at high
temperature predicted by the uncertainty principle.

We have provided simple fit formulas for the electrical and
thermal conductivities as well as the shear viscosity with a good
relative accuracy, which can be utilized in numerical simulations of
hydrodynamics of the quark plasma.

The present work can be expanded in a number of ways
  by extending the Lagrangian \eqref{eq:lagrangian} of the model.  The
  role of the confinement can be assessed by extending the NJL model
  to include the Polyakov loop at finite temperature. The mesons,
  which appear in the present model as scatterers, can carry momentum
  and charge in heavy-ion experiments and can contribute to
  transport, as established in numerical simulations of such
  experiments~(see
  Refs.~\cite{2008PhRvL.100q2301X,2013PhRvC..88d5204M,2014PhRvD..90k4009P,2012PhRvC..86e4902P}
  and references therein). In addition, having an access to the
  spectral functions of quarks, will allows us to compute the rates of
  the photon and dilepton emission from quark matter in the
  present model, which is again of interest for the description of
  heavy-ion collisions.

\section*{Acknowledgements}

 A.H.\ acknowledges support from the HGS-HIRe graduate
program at Frankfurt University. A.S.\ is supported by the Deutsche
Forschungsgemeinschaft (Grant No.\ SE 1836/3-2) and by the Helmholtz
International Center for FAIR. We acknowledge the
support by NewCompStar COST Action MP1304.

\appendix

\section{Details of the NJL-model calculations}
\label{app:B}

The constituent quark mass $m$ is found from the gap equation, 
which to leading order $\mathcal{O}(N_c^0)$ is given by the 
Hartree approximation  (see Fig.~\ref{fig:Hartree}) and analytically 
reads
\bea\label{eq:gap1}
S_0^{-1}=S^{-1}-G\langle\bar{\psi}\psi\rangle,
\eea
where $S_0^{-1}=\slashed p-m_0$, $S^{-1}=\slashed p-m$ are the free
and interacting quark propagators.  
The quark condensate $\langle\bar{\psi}\psi\rangle$, represented by 
the loop in diagram Fig.~\ref{fig:Hartree}, is given by 
\bea\label{eq:gap2}
\langle\bar{\psi}\psi\rangle = T\sum\limits_{m\in Z}
\int\frac{d\bm p}{(2\pi)^3}\Tr[S(\bm p,\omega_m)],
\eea
where the summation is over the fermionic Matsubara frequencies
$\omega_m=(2m+1)\pi T-i\mu$. The trace is over Dirac, color, and
flavor space and the quark propagator is given by
\be\label{eq:freeprop} 
S(\bm p,\omega_m)=\frac{\Lambda^+_p\gamma_0}
{i\omega_m-E_p}+
\frac{\Lambda^-_p\gamma_0}{i\omega_m+E_p},
\ee
where $\Lambda^+_p$ 
and $\Lambda^-_p$ are the projection operators onto 
positive and negative energy states
\be\label{eq:projectors}
\Lambda^\pm_p=\frac{E_p\gamma_0\mp\bm\gamma
\cdot\bm p\pm m}{2E_p}\gamma_0.
\ee
Substituting  Eqs.~(\ref{eq:freeprop}) and \eqref{eq:projectors}
into  Eq.~(\ref{eq:gap2})  and performing the Matsubara sums we 
obtain 
\bea\label{eq:gap4}
\langle\bar{\psi}\psi\rangle=-4N_cN_fmI_1,
\eea
where we defined
\bea\label{eq:integral1}
I_1=\frac{1}{4\pi^2}\int_0^\Lambda 
dp\frac{p^2}{E_p}[1-n^+(E_p)-n^-(E_p)].
\eea 
From Eqs.\ \eqref{eq:gap1} and \eqref{eq:gap4} follows
\bea\label{eq:mass1}
m=m_0+4GN_cN_fmI_1.
\eea
If $m_0\neq 0$, Eq.~\eqref{eq:mass1} always has a nontrivial solution
$m>m_0$. If $m_0=0$, there is a trivial solution $m=0$, but
Eq.~\eqref{eq:mass1} may have nontrivial solutions satisfying
\bea\label{eq:mass_chiral}
4GN_cN_fI_1=1.
\eea
In this case the vacuum energy is minimized by the solution with
the largest $m$~\cite{2005PhR...407..205B}.  At high densities and
temperatures ($T>T_c\simeq 190$ MeV for $\mu=0$ or
$\mu>\mu_c\simeq 332$ MeV for $T=0$) Eq.~\eqref{eq:mass_chiral} does
not have solutions anymore, and we find that chiral symmetry is restored
with $m_0=m=0$.

The meson propagators are obtained  from the Bethe-Salpeter equation,
shown in Fig.~\ref{fig:BS_eq}, 
\bea\label{eq:meson_prop}
D_M=G+G\Pi_{M}D_M=\frac{G}{1-G\Pi_M},
\eea
where the quark-antiquark polarizations for the $\sigma$-meson and pion
$\Pi_M, M=\sigma,\pi$ are given by the formula
\bea\label{eq:Pi_M}
\Pi_M(\bm p, \omega_n)&=&-T\sum\limits_{m\in Z}
\int\frac{d\bm q}{(2\pi)^3}\Tr[\Gamma_M
S(\bm q+\bm p,\omega_m+\omega_n)\nonumber\\
&\times&\Gamma_M S(\bm q,\omega_m)]
\eea
with $\Gamma_\sigma=1$, $\Gamma_\pi=i\gamma_5\tau_j$, $j=1,2,3$.  
From Eq.\ \eqref{eq:freeprop} we obtain
\bea\label{eq:Pi_M1}
\Pi_M(\bm p, \omega_n)
=-\int\frac{d\bm q}{(2\pi)^3}
\sum\limits_{\pm\pm}{\cal T}^{\pm\pm}_M{\cal S}^{\pm\pm},
\eea
where we defined
\bea\label{eq:T_M}
 {\cal T}^{\pm\pm}_M&=&\Tr[\Gamma_M\Lambda^{\pm}_{q+p}
\gamma_0\Gamma_M\Lambda^\pm_q\gamma_0],\\
\label{eq:matsubarasums_M}
{\cal S}^{\pm\pm} &=& T\sum\limits_{m}\frac{1}
{(i\omega_m +i\omega_n- E^\pm_{q+p})
(i\omega_m- E^\pm_q)},\nonumber\\
\eea 
with $E_q^\pm =\pm E_q$. In Eq.~\eqref{eq:Pi_M1} the sum runs
over all possible combinations of the signs of $E_{q+p}^\pm$ and
$E_q^\pm$. It is enough to calculate only one term of the sum, for
example $\cal T^{++}_M$ and $\cal S^{++}$, and the others can be
obtained by an appropriate choice of signs. Below we skip these signs and
recover them in the final expressions.  The computation of traces
gives 
\bea\label{eq:trace_M2}{\cal T}_{M} =N_cN_f
\frac{P_Mm^2+E_{q+p}E_q-\bm q \cdot (\bm q+\bm p)}{E_{q+p}E_q}, 
\eea 
where $P_\sigma=1,$ and $P_\pi=-1$.

The sum over the Matsubara frequencies gives
\bea\label{eq:matsubara_M}
{\cal S}=\frac{n^+(E_q)-n^+(E_{q+p})}
{E_q-E_{q+p}+i\omega_n}.
\eea

\begin{widetext}
Using the results \eqref{eq:trace_M2} and \eqref{eq:matsubara_M} we find for the
polarization tensor \eqref{eq:Pi_M1} 
\bea\label{eq:Pi_M2}
\Pi_M(\bm p, \omega_n)=
-N_cN_f\int\frac{d\bm q}{(2\pi)^3}
\bigg\{\frac{P_Mm^2+E_{q+p}E_q-
\bm q\cdot(\bm q+\bm p)}{E_{q+p}E_q}\bigg[
\frac{n^+(E_q)-n^+(E_{q+p})}
{E_-+i\omega_n}
-\frac{n^-(E_q)-n^-(E_{q+p})}
{-E_-+i\omega_n}\bigg]\nonumber\\
+\frac{P_Mm^2-E_{q+p}E_q-
\bm q\cdot(\bm q+\bm p)}{E_{q+p}E_q}\bigg[
\frac{n^-(E_q)+n^+(E_{q+p})-1}
{-E_++i\omega_n}
-\frac{n^+(E_q)+n^-(E_{q+p})-1}
{E_++i\omega_n}\bigg]\bigg\},\nonumber\\
\eea
with $E_\pm=E_q\pm E_{q+p}$. 
Define the short-hand notation 
$
N_M(\bm p,\omega_n)=-(\bm p^2+\omega_n^2)
-2(1+P_M)m^2,
$
which gives for the $\pi$- and $\sigma$-modes, respectively,
$N_\pi(\bm p,\omega_n)=-(\bm p^2+\omega_n^2)$ and 
$ N_\sigma(\bm p,\omega_n) =N_\pi(\bm p,\omega_n)-4m^2.$
Then Eq.\ \eqref{eq:Pi_M2} can be written in a compact form 
\bea\label{eq:Pi_M5}
\Pi_M(\bm p, \omega_n)=2N_cN_f  \big[ 2I_1+
N_M(\bm p,\omega_n)I_2(\bm p, \omega_n)\big],
\eea
where 
\bea\label{eq:integral_2}
I_2(\bm p,\omega_n)&=&
\int\frac{d\bm q}{(2\pi)^3}
\frac{1}{4E_{q}E_{q-p}}
\bigg\{\frac{2E_+}{\omega_n^2+E_+^2}+\nonumber\\
&+&\frac{(i\omega_n-E_+)[n^-(E_{q})+n^+(E_{q-p})]-(i\omega_n+E_+)[n^+(E_{q})+n^-(E_{q-p})]}
{\omega_n^2+E_+^2}\nonumber\\
&+&\frac{(i\omega_n+E_-)[n^+(E_{q})-n^+(E_{q-p})]-(i\omega_n-E_-)[n^-(E_{q})-n^-(E_{q-p})]}
{\omega_n^2+E_-^2}\bigg\},
\eea 
with redefined $E_\pm=E_q\pm E_{q-p}$. 
\end{widetext}
Now we can write the meson propagator according to
Eqs.~\eqref{eq:meson_prop} and \eqref{eq:Pi_M5} as
\bea\label{eq:meson_prop1}
D_M^{-1}(\bm p, \omega_n)&=&G^{-1}-2N_cN_f 
\nonumber\\
&\times &[2I_1 +N_M(\bm p,\omega_n)I_2(\bm p, \omega_n)].
\eea
The momentum-independent meson mass is defined 
as the pole of the propagator in real space-time
for $\bm p=0$ ($i\omega_n\to m_M+i\delta$)
\bea\label{eq:meson_mass}
{\rm Re}D_M^{-1}[\bm 0,-i(m_M+i\delta)]=0,
\eea
or, 
\bea\label{eq:meson_mass1}
m_\pi^2I_2(\bm 0,-im_\pi)&=&\frac{1-4GN_cN_fI_1}
{2GN_cN_f},\\
\label{eq:meson_mass2}
(m_\sigma^2 -4m^2)I_2(\bm 0,-im_\sigma)
&=&\frac{1-4GN_cN_fI_1}{2GN_cN_f},
\eea
where we took into account that
${\rm Re}I_2[\bm 0,-i(m_M+i\delta)]=I_2(\bm 0,-im_M)$ in principal-value prescription.  
From Eq.~\eqref{eq:integral_2} we have for $\bm p=0$ ($E_-=0$,
$E_+=2E_q$)
\bea\label{eq:integral2_p0}
I_2(\bm 0,\omega_n)
=\frac{1}{8\pi^2}\int_0^\Lambda\!\! q^2dq
\frac{1-n^+(E_{q})-n^-(E_{q})}
{E_q(E_q^2+\omega_n^2/4)},
\eea
and after analytic continuation $i\omega_n\to\omega+i\delta$ we obtain
\bea\label{eq:re_integral2_p0}
{\rm Re}I_2(\bm 0,\omega)&=&I_2(\bm 0,-i\omega)\nonumber\\
&=&
\frac{1}{8\pi^2}\int_m^{E_\Lambda} \!\! qdE_q
\frac{1-n^+(E_{q})-n^-(E_{q})}
{E_q^2-\omega^2/4},\nonumber\\
\eea
where  $E_\Lambda=\sqrt{m^2+\Lambda^2}$.
For $m<\omega/2$ the integrand of Eq.~\eqref{eq:re_integral2_p0}
has a single pole, and the integral should be understood in the sense
of its principal value. 
For $\omega>0$ we have
\bea\label{eq:im_integral2_p0}
{\rm Im}I_2(\bm 0,\omega)=
&&\frac{\sqrt{\omega^2 -4m^2}}{16\pi\omega}\nonumber\\
&&\hspace{-1.8cm}\times\frac{\sinh(\beta\omega/2) \theta(\omega-2m)}{\sinh(\beta\omega/2)+\cosh(\beta\mu)}
\theta(2E_\Lambda -\omega).
\eea
As seen from Eq.~\eqref{eq:im_integral2_p0}, the pion propagator
obtains an imaginary part when $\omega>2m$, therefore it becomes
unstable to the on-shell decay into a quark-antiquark pair.

From Eqs.~\eqref{eq:integral1}, \eqref{eq:mass1}, \eqref{eq:meson_mass1},
\eqref{eq:meson_mass2}, and \eqref{eq:re_integral2_p0} we obtain the following equation for $m_\pi$ and $m_\sigma$
\bea\label{eq:meson_masses}
1 &=&\frac{GN_cN_f}{\pi^2}\int_m^{E_\Lambda} \!\!\! qdE(E^2-\alpha_M
m^2)\frac{1-n^+(q)-n^-(q)}{E^2-(m_M/2)^2},
\nonumber\\
\eea
where $\alpha_\pi=0$, $\alpha_\sigma=1$.  The solutions provided by
Eq.~\eqref{eq:meson_masses} are displayed in Fig.~\ref{fig:masses} and
are discussed in the main text.

The Mott temperature $T_{\rm M}$, defined by the condition $m_\pi=2m$, can be found from the following equation
\bea\label{eq:Mott_temp}
&&\frac{GN_cN_f}{\pi^2}\int_0^{\Lambda}
dqE[1-n^+(q)-n^-(q)]=1. 
\eea 

\begin{figure}[tb] 
\begin{center}
\includegraphics[width=8.4cm,keepaspectratio]{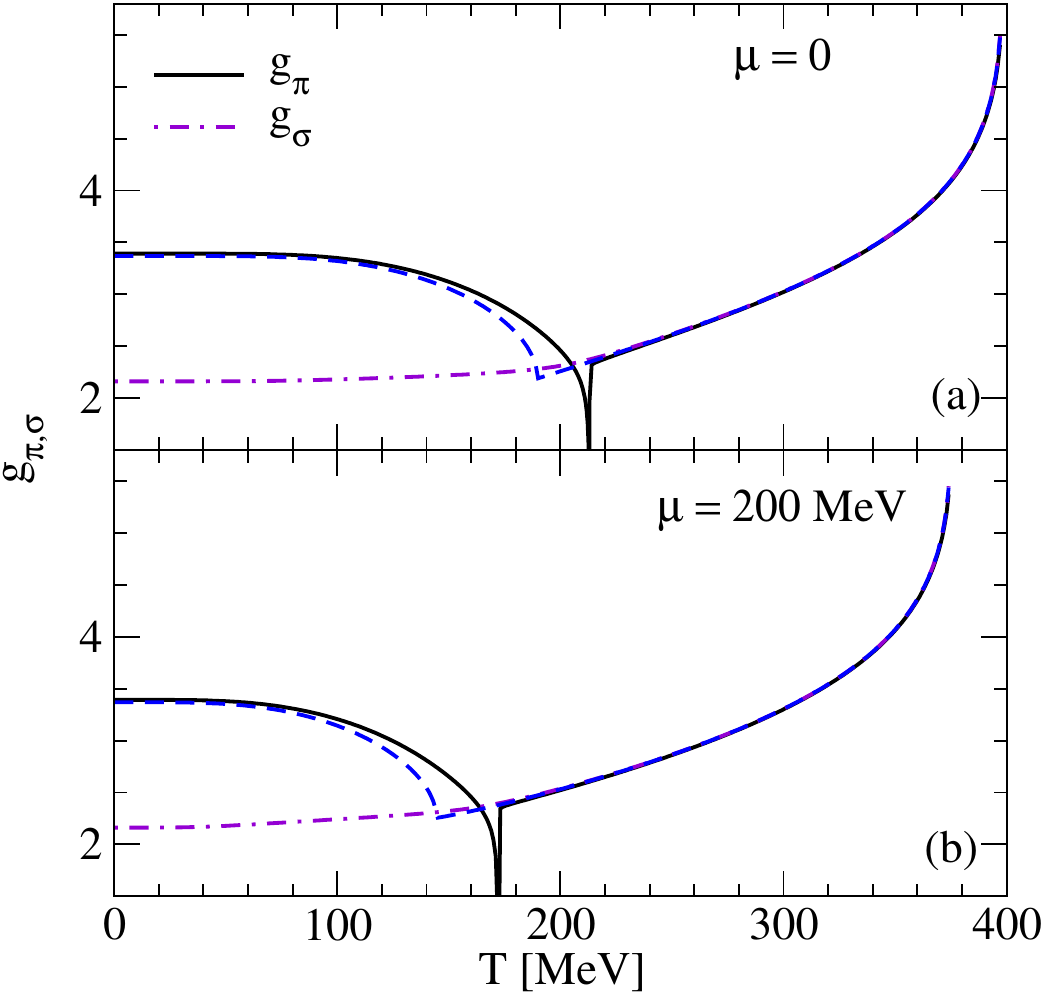}
\caption{ Dependence of the couplings $g_{\pi}$ and $g_{\sigma}$ on
  temperature for zero and nonzero chemical potentials. The chiral
  limit for the $\pi$-meson is shown by short-dashed lines. }
\label{fig:couplings} 
\end{center}
\end{figure}

Now we evaluate the meson propagator using the standard mass-pole
approximation (the imaginary part of the pion self-energy is
neglected)
\bea\label{eq:meson_prop2}
D_M(\bm p, -i\omega)=\frac{-g^2_M}
{\omega^2-\bm p^2-m_M^2+i\varepsilon},
\eea
where the quark-meson coupling is defined as the residue
of the full meson propagator at vanishing momentum
\bea\label{eq:coupling}
g^{-2}_M=-\frac{d}{d\omega^2}D^{-1}_M[\bm 0,-i\omega]\bigg\vert_{\omega^2=m_M^2}.
\eea
Employing Eqs.\ \eqref{eq:meson_prop1} and \eqref{eq:integral2_p0} we
obtain for the $\pi$- and $\sigma$-modes
\bea\label{eq:coupling_pi}
g^{-2}_\pi &=&
2N_cN_f\bigg[I_2(\bm 0, -im_\pi)\nonumber\\
&+&m_\pi^2
\frac{d}{d\omega^2}I_2(\bm 0,
-i\omega)\bigg\vert_{\omega^2=m_\pi^2}\bigg],\\
\label{eq:coupling_sigma}
g^{-2}_\sigma &=&
2N_cN_f\big[I_2(\bm 0, -im_\sigma)\nonumber\\
&+&(m_\sigma^2-4m^2)
\frac{d}{d\omega^2}I_2(\bm 0, -i\omega)
\bigg\vert_{\omega^2=m_\sigma^2}\bigg].
\eea
To compute the derivative appearing in
Eqs.~\eqref{eq:coupling_pi}-\eqref{eq:coupling_sigma} 
one can formally replace
${d}/{d\omega^2}\to -\frac{1}{4}({d}/{dE_q^2})$
in Eq.~\eqref{eq:re_integral2_p0} and integrate by
parts to obtain 
\bea\label{eq:integral2_deriv}
64\pi^2\frac{d}{d\omega^2}I_2(\bm 0,-i\omega)
&=&-
\frac{\Lambda[1-n^+(E_{\Lambda})-n^-(E_{\Lambda})]}
{E_\Lambda(E_\Lambda^2-\omega^2/4)}\nonumber\\
&&\hspace{-3.5cm}+
\int_m^{E_\Lambda} dE_q\frac{1}{q(E_q^2-\omega^2/4)}
\bigg\{\frac{m^2}{E_q^2}
[1-n^+(E_{q})-n^-(E_{q})]\nonumber\\
&&\hspace{-3.5cm}+\frac{q^2}{E_qT}[n^+(E_{q})(1-n^+(E_{q}))
+n^-(E_{q})(1-n^-(E_{q}))]\bigg\}.\nonumber\\
\eea
Figure \ref{fig:couplings} shows the temperature dependence of the
couplings at zero and nonzero chemical
potentials. Note that the jump in the coupling of the $\pi$-meson
arises at the Mott temperature, \ie, $g_\pi\to 0$ at $T\to T_{\rm M}$,
which can be verified from Eq.~\eqref{eq:coupling_pi}, where the
integral \eqref{eq:integral2_deriv} diverges for $\omega\to 2m$. The
two couplings are almost identical above the Mott temperature.

In the chiral limit the overall behavior of the coupling constants
remains the same, except for the absence of a discontinuity in the
$\pi$-meson coupling at the Mott line.  Above the Mott temperature the
$\pi$- and $\sigma$-meson coupling constants nearly coincide. 
\begin{figure*}[t] 
\begin{center}
\includegraphics[width=12.cm,keepaspectratio]{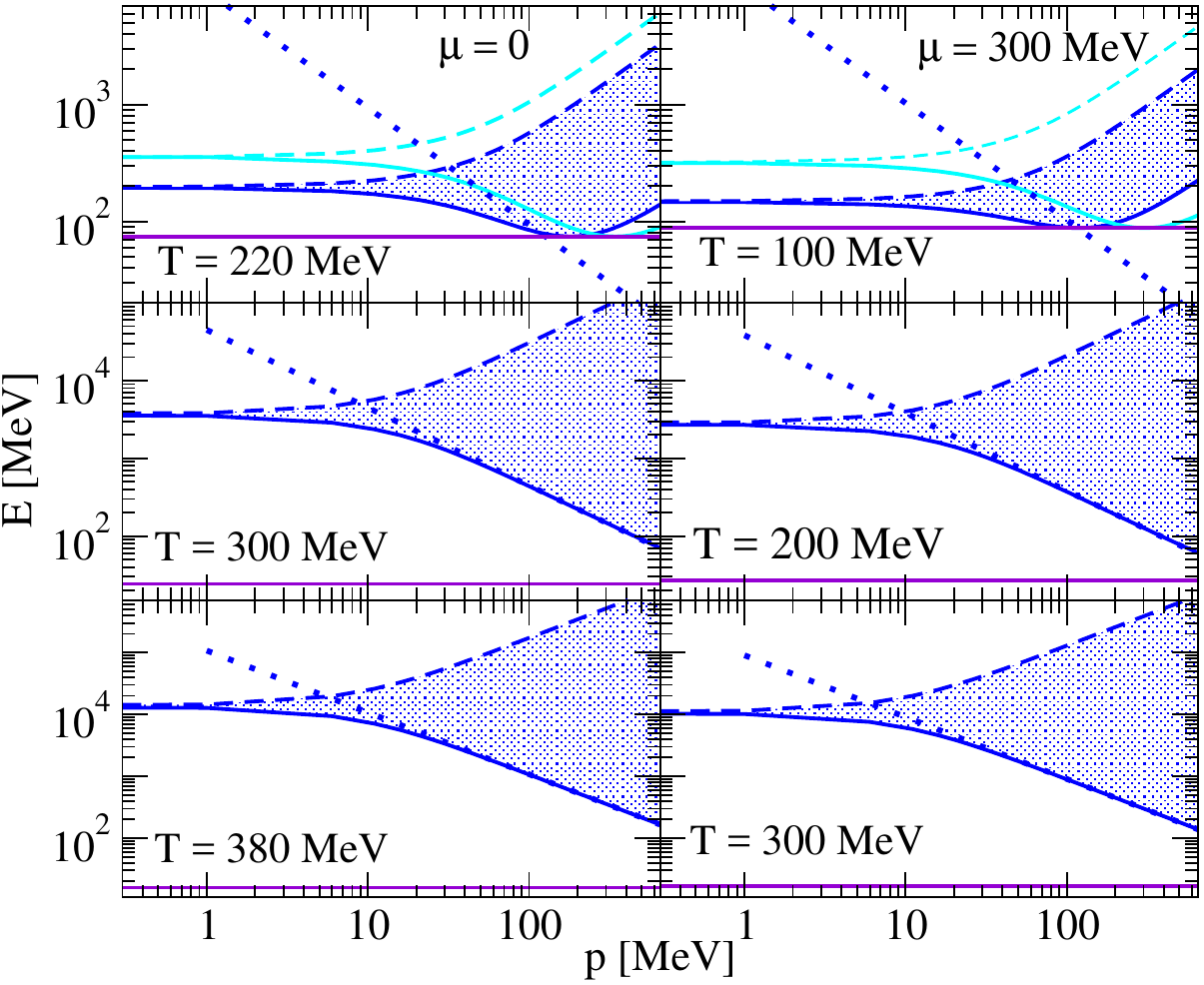}
\caption{Dependence of the ranges of integration (shaded area) in the
  self-energies \eqref{eq:im_self} and \eqref{eq:im_self_anti} on
  momentum at various values of temperature and chemical
  potential. The ranges are limited by $E_{{\rm min},\pi}$ (heavy
  solid line) and $E_{{\rm max},\pi}$ (heavy dashed line) for
  $\pi$-mesons and by $E_{{\rm min},\sigma}$ (light solid line) and
  $E_{{\rm max},\sigma}$ (light dashed line).  The chiral limit
  Eq.~\eqref{eq:E_min_max_chiral} is shown by the dotted lines. The value
  of the quark mass is shown by the heavy horizontal line.}
\label{fig:limits_self} 
\end{center}
\end{figure*}

\section{Meson-exchange quark self-energy}
\label{app:C}

Here we provide some details of the computation of the self-energy
\eqref{eq:self} and its imaginary part. Analogous calculations for
relativistic nucleonic matter were carried out in
Ref.~\cite{2013PhRvC..88a5209C} and for two-flavor quark matter in
Ref.~\cite{LKW15}. Substituting the quark and meson propagators into
Eq.~\eqref{eq:self}, see Eqs.~\eqref{eq:freeprop},
\eqref{eq:projectors}, \eqref{eq:meson_prop2}, we find
\bea\label{eq:self1}
\Sigma^M(\bm p,\omega_n)
&=&g^2_M\int \frac{d\bm q}{(2\pi)^32} \sum\limits_{\pm}\frac{1}{2E_q^\pm}
\nonumber\\
&&\hspace{-1.5cm} \times
T\sum\limits_{m}\frac
{\Gamma_M (E_q^\pm\gamma_0-\bm\gamma \cdot \bm q+ m)\Gamma_M}{(i\omega_m-E^\pm_q)
[(\omega_n-\omega_m)^2+E_M^2]},
\eea
where $E_M^2=(\bm p-\bm q)^2+m_M^2$.  For $\Gamma_\sigma=1$ and
$\Gamma_\pi=i\gamma_5\tau_j$ we can write
$\Gamma_M (E_q^\pm\gamma_0- \bm\gamma \cdot \bm q+ m)\Gamma_M=
E_q^\pm\gamma_0-\bm\gamma\cdot \bm q+P_Mm$,
with $P_\sigma=1$, $P_\pi=-1$, therefore
\bea\label{eq:self2}
\Sigma^M(\bm p,\omega_n)&=&
P_Mm\Sigma^M_s+i\omega_n\gamma_0\Sigma^M_0
-\bm p\cdot \bm\gamma\Sigma^M_v, \quad
\eea
where we defined 
\bea\label{eq:self_s}
\Sigma^M_s &=&
g^2_M\int \frac{d\bm q}{(2\pi)^3} 
\sum\limits_{\pm}\frac{\mathscr{S}^{\pm}}{2E_q^\pm},\\
\label{eq:self_0}
\Sigma^M_0 &=&
g^2_M\int \frac{d\bm q}{(2\pi)^3  } 
\frac{1}{2i\omega_n}\sum\limits_{\pm}\mathscr{S}^{\pm},\\
\label{eq:self_v}
\Sigma^M_v &=&
g^2_M\int \frac{d\bm q}{(2\pi)^3}
\frac{\bm q\cdot\bm p}{p^2}
\sum\limits_{\pm}\frac{\mathscr{S}^{\pm}}{2E_q^\pm},
\eea
and 
\bea\label{eq:matsubarasums1}
\mathscr{S}^{\pm}
&=&\frac{1}{2E_M}\bigg[\frac{n^+(E^\pm_q)+n_B(-E_M)}
{E^\pm_q+E_M-i\omega_n}\nonumber\\
&&\hspace{0.8cm}-\frac{n^+(E^\pm_q)+n_B(E_M)}
{E^\pm_q-E_M-i\omega_n}\bigg].
\eea
Defining $E_\pm =E_q\pm E_M$ and
 using the properties $n_B(-E)=-1-n_B(E)$, 
$n^+(-E)=1-n^-(E)$ we obtain
\bea\label{eq:matsubarasums2}
\sum\limits_{\pm}\mathscr{S}^{\pm}
&=&\frac{1}{2E_M}\bigg[\frac{E_+\mathscr{C}_3-2i\omega_n \mathscr{C}_1}
{E_+^2+\omega_n^2}\nonumber\\
&&\hspace{0.8cm}-
\frac{E_-\mathscr{C}_3+2i\omega_n \mathscr{C}_2}
{E_-^2+\omega_n^2}\bigg],
\eea
where $\mathscr{C}_1$, $\mathscr{C}_2$, and $\mathscr{C}_3$ are defined 
in Eqs.~\eqref{eq:Z_123}.
In the same manner we obtain
\bea\label{eq:matsubarasums3}
\sum\limits_{\pm}\frac{\mathscr{S}^{\pm}}{2E_q^\pm}
&=&\frac{1}{4E_qE_M}\bigg[\frac{i\omega_n \mathscr{C}_3-2E_+ \mathscr{C}_1}
{E_+^2+\omega_n^2}\nonumber\\
&&\hspace{1.2cm}-
\frac{i\omega_n \mathscr{C}_3+2E_- \mathscr{C}_2}
{E_-^2+\omega_n^2}\bigg].
\eea
Now using Eqs.~\eqref{eq:matsubarasums2} and \eqref{eq:matsubarasums3}
in Eqs.~\eqref{eq:self_s}-\eqref{eq:self_v} we obtain 
Eqs.~\eqref{eq:self_sv1} and \eqref{eq:self_01} of the main text. 

We now turn to the computation of the imaginary parts of
Eqs.~\eqref{eq:self_sv1} and \eqref{eq:self_01}, which are extracted using
the Dirac identity by writing
\bea\label{eq:dirac1}
&&{\rm Im}\frac{1}{E^2+\omega_n^2}\bigg\vert_{i\omega_n
\to p_0+i\varepsilon}
=\pi{\rm sgn} (p_0)\delta(E^2-p_0^2)\nonumber\\
&&\hspace{2cm}=
\frac{\pi}{2p_0}[\delta(E+p_0)+\delta(E-p_0)].
\eea
For on-shell quarks with $p_0=\sqrt{\bm p^2+m^2}$
Eqs.~\eqref{eq:self_sv1} and \eqref{eq:self_01} generate  four
contributions after 
analytical continuation and application of the Dirac identity, 
each of which is proportional to a $\delta$-function
\bea\label{eq:deltas}
\delta(E_M+E_q+p_0)\equiv \delta_1,\,\,
\delta(E_M+E_q-p_0)\equiv \delta_2,\\
\delta(E_M-E_q+p_0)\equiv \delta_3,\,\,
\delta(E_M-E_q-p_0)\equiv \delta_4.
\eea
For $p_0 \geq 0$, the term $\delta_1$ vanishes trivially. As we will
show later, the terms $\delta_2$ and $\delta_3$ do not give 
any contribution as well. 
For $\delta_4$ we can write 
\bea\label{eq:dirac2} 
\delta_4
=\frac{E_M}{pq}\delta(x-x_-),\,\,\,
x_-=\frac{m_M^2-2m^2-2E_qp_0}{2pq}, \quad
\eea 
and for an arbitrary function $f(\bm q)$ we then have
\bea\label{eq:dirac3} 
\int \frac{d\bm
  q}{(2\pi)^3}f(\bm q)\delta_4 = \int_{E_{\rm min}}^{E_{\rm
    max}}\frac{d E_q}{(2\pi)^2}\frac{E_qE_M}{p}f(q,x_-), \quad 
\eea 
where the limits of integration are found from the condition
$x_-^2\le 1$
\bea\label{eq:cond_x-}
\left(\frac{m_M^2-2m^2-2E_qp_0}{2pq}\right)^2\le 1, 
\eea 
which leads us to Eq.\ \eqref{eq:E_min_max} of the main text.

In analogy to Eqs.~\eqref{eq:dirac2} and \eqref{eq:dirac3} we have
\bea\label{eq:dirac4}
\delta_2 +\delta_3 &=&\delta(E_M+E_q-p_0)
+{\delta(E_M-E_q+p_0)}\nonumber\\
&=&\frac{E_M}{pq}\delta(x-x_+),\quad
x_+=\frac{m_M^2-2m^2+2E_qp_0}{2pq},\nonumber\\
\eea
therefore
\bea\label{eq:dirac5}
&&\int \frac{d\bm q}{(2\pi)^3}f(\bm q)
(\delta_2 +\delta_3)\nonumber\\
&&\hspace{0.5cm}=
\int_m^\infty \frac{d E_q}{(2\pi)^2}
\frac{E_qE_M}{p}f(q,x_+)\theta(1-x_+^2).
\eea
The condition $x_+^2\le 1$ is satisfied for
$E_q\in (E'_{\rm min},E'_{\rm max})$ with $E'_{\rm min}=-E_{\rm max}$,
$E'_{\rm max}=-E_{\rm min}$, as can be seen from
Eq.~\eqref{eq:E_min_max} (note that $x_+$ is obtained from $x_-$ by the
inversion $p_0\to -p_0$).  In this case the integration range is
empty, and the integral vanishes.

We are now in a position to take the imaginary parts of the
self-energies; keeping the only nonvanishing part $\propto {\rm Im}
[E_-^2+\omega_n^2]^{-1}=\pi (2p_0)^{-1}\delta_4$ we find
\bea\label{eq:im_self_sv}
{\rm Im}\Sigma^M_{s,v} &=&-
g^2_M\int \frac{d\bm q}{(2\pi)^3}
\mathscr{Q}_{s,v}(\bm q)\frac{p_0 \mathscr{C}_3
+2E_- \mathscr{C}_2}{4E_qE_M}
\frac{1}{E_-^2+\omega_n^2}\nonumber\\
&&\hspace{-1.5cm}=\frac{g^2_M}{16\pi p}
\int_{E_{\rm min}}^{E_{\rm max}}d E_q
\mathscr{Q}_{s,v}(q,x_-)[n_B(E_M)+n^-(E_q)], \nonumber\\
\eea
and
\bea\label{eq:im_self_0}
{\rm Im}\Sigma^M_0 &=&
g^2_M\int \frac{d\bm q}{(2\pi)^3} 
\mathscr{Q}_0(q)\frac{2p_0 \mathscr{C}_2+E_-\mathscr{C}_3}{4 E_qE_M}
\frac{1}{E_-^2+\omega_n^2}\nonumber\\
&=&
\frac{g^2_M}{16\pi p}
\int_{E_{\rm min}}^{E_{\rm max}}d E_q
\mathscr{Q}_0 (q)[n_B(E_M)+n^-(E_q)]. \nonumber\\
\eea
Combining Eqs.~\eqref{eq:im_self_sv} and \eqref{eq:im_self_0} we finally
obtain Eq.~\eqref{eq:im_self} of the main text.  The imaginary part of the
on-shell self-energy of antiquarks ($p_0=-E_p$),
given by 
Eq.~\eqref{eq:im_self_anti}, is obtained through a
similar calculation. To support the discussion of the various
contributions to the self-energies \eqref{eq:im_self} and
\eqref{eq:im_self_anti} in the main text we show in
Fig.~\ref{fig:limits_self} the ranges of integration in these
equations.

\section{Thermodynamic quantities}
\label{app:D}

The particle number and entropy densities of quark matter within the $1/N_c$ approximation are
given by the formulas
\bea\label{eq:quark_number_density}
n&=& \frac{N_cN_f}{\pi^2}\int_0^\infty 
p^2dp[n^+(E_p)-n^-(E_p)],\\
\label{eq:entropy}
s&=&\frac{N_cN_f}{\pi^2}\!\!\int_0^\infty\!\!\! p^2dp
[\beta(E_p-\mu)n^+(E_p) +\beta(E_p+\mu)n^-(E_p) \nonumber\\
&&-\log(1-n^+(E_p))-\log(1-n^-(E_p))],
\eea
with $n^\pm(E)=[e^{\beta(E\mp\mu)}+1]^{-1}$.
The integrals in Eqs.~\eqref{eq:quark_number_density} and \eqref{eq:entropy}
should be calculated without momentum cutoff, but
for momenta $p>\Lambda$ the quark energy should be evaluated
with its bare mass, \ie, $E_p=\sqrt{p^2+m_0^2}$.
\begin{figure}[!] 
\begin{center}
\includegraphics[width=8.0cm,keepaspectratio]{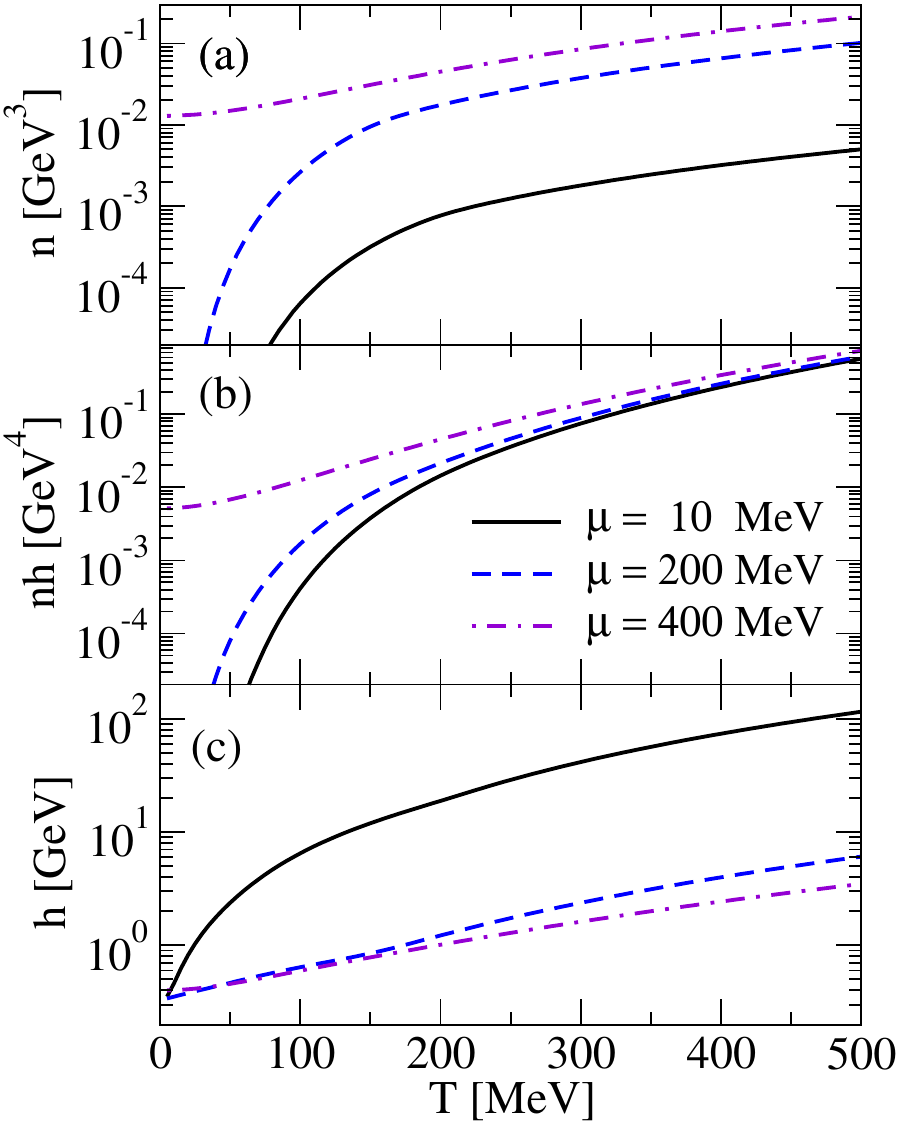}
\caption{ Dependence of (a) quark number density, (b)  enthalpy density, and (c) the enthalpy per particle on the temperature for several values of the chemical
  potential.}
\label{fig:enthalpy} 
\end{center}
\end{figure}
\begin{figure}[!] 
\begin{center}
\includegraphics[width=8.0cm,keepaspectratio]{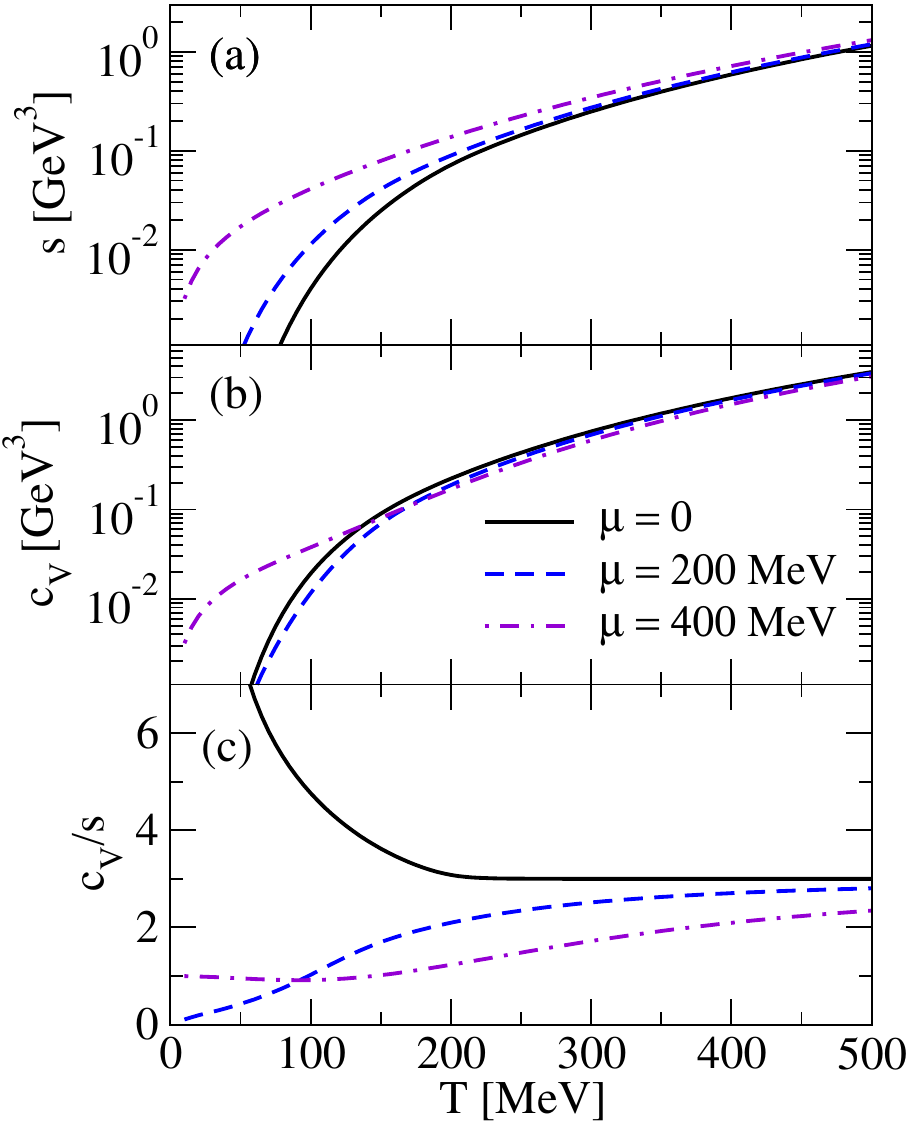}
\caption{ Dependence of (a) entropy density, (b) specific heat, and (c)
  their ratio on the temperature for several values of the chemical
  potential.}
\label{fig:entropy_heat} 
\end{center}
\end{figure}

The enthalpy per particle is defined as
\bea\label{eq:enthalpy}
h&=& \frac{Ts+\mu n}{n}=
\frac{N_cN_f}{\pi^2 n}\int_0^\infty p^2dp\nonumber\\
&\times &\left(E_p+\frac{p^2}{3E_p}\right)[n^+(E_p)+n^-(E_p)].
\eea

Figure~\ref{fig:enthalpy} shows the quark number density, the enthalpy
density, and the enthalpy per particle given by
Eqs.~\eqref{eq:quark_number_density} and \eqref{eq:enthalpy}. In the
nondegenerate regime $T\gg \mu$ we distinguish the following limiting
cases: (i) $T\gg m$, \ie, high temperatures where chiral symmetry is
(approximately) restored; and (ii) $T\ll m$, \ie, the regime where quarks
are nonrelativistic.  In the first case $p\simeq \varepsilon\sim T$,
therefore we have the scalings $n\propto \mu T^2$, $nh\propto T^4$, and
$h\propto T^2/\mu\to \infty$ at $\mu\to 0$. In the second case the
integrands of Eqs.~\eqref{eq:quark_number_density} and
\eqref{eq:enthalpy} are exponentially suppressed by the distribution function for energies
$\varepsilon-m\gtrsim T$, and we find the
scalings $n\propto m\mu(mT)^{1/2}e^{-m/T}$,
$nh\propto m(mT)^{3/2}e^{-m/T}$, and $h\simeq mT/\mu\gg m$. Thus, the
quark number density and the enthalpy density are exponentially
suppressed because of the nonvanishing quark condensate at low
temperatures.  The enthalpy per particle again diverges as the
chemical potential tends to zero.  In the opposite, strongly degenerate
limit $T\ll\mu$ consider the cases: (i) $\mu\gg m$ and (ii)
$\Delta\gg T$, where $\Delta \equiv m-\mu$.  The first case is
realized at high chemical potentials $\mu\gtrsim 350$ MeV, where we
have degenerate matter along with (approximate) chiral symmetry restoration. In
this case all three quantities depend only on the chemical potential:
$n\propto \mu^3$, $nh\propto \mu^4$, and $h\to\mu$.  The second case is
realized for intermediate values of the chemical potential
$\mu\lesssim 300$ MeV, where the constituent quark mass still exceeds
the chemical potential. In this case the quark number density and the
enthalpy density vanish exponentially when $T\to 0$ according to the
scalings $n\propto (mT)^{3/2}e^{-\Delta/T}$,
$nh\propto m(mT)^{3/2}e^{-\Delta/T}$, with $\Delta>0$, and the
enthalpy per particle has a finite limit $h\to m$, as seen in
Fig.~\ref{fig:enthalpy}.

Using Eqs.~\eqref{eq:quark_number_density}-\eqref{eq:entropy} we can calculate the specific heat of
quark matter via the standard formula for the heat capacity per unit volume
\bea\label{eq:c_v}
c_V=T\left(\frac{\partial s}{\partial T}\right)_{n}=
-\beta\left(\frac{\partial s}{\partial\beta}\right)_{\mu}
-\beta\left(\frac{\partial s}{\partial\mu}\right)_{\beta}
\left(\frac{\partial\mu}{\partial\beta}\right)_{n}\!\!.\quad
\eea
Using the relations
\bea\label{eq:fermi_deriv}
\left(\frac{\partial n^\pm}{\partial\beta}\right)_{\mu}&=&
-(E_p\mp \mu) n^\pm(1-n^\pm),\\
\label{eq:fermi_deriv1}
\left(\frac{\partial n^\pm}{\partial\mu}\right)_{\beta}&=&
\pm \beta n^\pm(1-n^\pm),
\eea
by taking the derivatives of Eq.~\eqref{eq:entropy}  we find
\bea\label{eq:entropy1}
\left(\frac{\partial s}{\partial\beta}\right)_{\mu}
&=&-\frac{N_cN_f}{\pi^2 T}\int_0^\infty 
p^2dp[(E_p-\mu)^2n^+(1-n^+)\nonumber\\
&+&
(E_p+\mu)^2n^-(1-n^-)],\\
\nonumber\\
\label{eq:entropy2}
\left(\frac{\partial s}{\partial\mu}\right)_{\beta}
&=&\frac{N_cN_f}{\pi^2 T^2}\int_0^\infty p^2dp
[(E_p-\mu)n^+(1-n^+)\nonumber\\
&-&(E_p+\mu)n^-(1-n^-)],
\eea
which implies that  Eq.~\eqref{eq:c_v} can be written as
\bea\label{eq:c_v1}
c_V
&=&
\frac{N_cN_f}{\pi^2 T^2}\int_0^\infty p^2dp
[(E_p-\mu)(E_p-\mu^*)n^+(1-n^+)\nonumber\\
&+&(E_p+\mu)(E_p+\mu^*)n^-(1-n^-)],
\eea
where we introduced
\bea\label{eq:mu_star}
\mu^*(\beta,\mu)=
\mu+\beta\left(\frac{\partial\mu}{\partial\beta}\right)_{n}.
\eea
Here we neglected the dependence of the constituent quark mass on the
temperature and the chemical potential, as it is of minor importance
above the Mott temperature. In order to find
$\left(\partial\mu/\partial\beta\right)_{n}$ we take the derivative of
Eq.~\eqref{eq:quark_number_density} with respect to $\beta$ for
$n={\rm const}$.  As the left-hand side vanishes, we obtain
\bea\label{eq:quark_number_deriv}
&&\hspace{-0.5cm}\int_0^\infty p^2dp\left[
-(E_p-\mu) n^+(1-n^+)+(E_p+\mu) n^-(1-n^-)\right]\nonumber\\
&&\hspace{-0.1cm}+\beta\left(\frac{\partial\mu}{\partial\beta}\right)_{n}
\int_0^\infty p^2dp\left[ n^+(1-n^+)+ n^-(1-n^-)\right]
=0.\nonumber\\
\eea
From Eqs.~\eqref{eq:mu_star} and
\eqref{eq:quark_number_deriv} it follows that
\bea\label{eq:mu_star_odd}
\mu^*(\beta,-\mu)=-\mu^*(\beta,\mu)
\eea
and 
\bea\label{eq:quark_number_deriv1}
&&\int_0^\infty \!\!\! p^2dp\bigg[(E_p-\mu^*) 
n^+(1-n^+)\nonumber\\
&&\hspace{1cm}-(E_p+\mu^*) n^-(1-n^-)\bigg]=0,
\eea
therefore  Eq.~\eqref{eq:c_v1} can be written as 
\bea\label{eq:c_v2}
c_V&=&\frac{N_cN_f}{\pi^2 T^2}\int_0^\infty
p^2dpE_p[(E_p-\mu^*)n^+(1-n^+) \nonumber\\
&+&(E_p+\mu^*)n^-(1-n^-)],
\eea
which is an even function of $\mu$, as expected.

Figure~\ref{fig:entropy_heat} shows the entropy density, the specific
heat capacity, and their ratio given by Eqs.~\eqref{eq:entropy} and
\eqref{eq:c_v2}.  In the nondegenerate ultrarelativistic limit
$\mu^*\simeq 3\mu\ll T$, $s$ and $c_V$ scale as $T^3$, therefore their
ratio tends to its classical limit $c_V/s \to 3$. In the
nondegenerate nonrelativistic limit ($m\gg T\gg \mu$) we find
$\mu^*\propto \mu m/T\ll m$ as well as the scalings
$s\propto m^2(mT)^{1/2}e^{-m/T}$ and
$c_V\propto m^3(m/T)^{1/2}e^{-m/T}$, which demonstrate the exponential
suppression of these quantities by the quark condensate. In this case
their ratio $c_V/s\simeq m/T$ diverges as $T\to 0$, as seen from the
solid line in panel (c). At very high chemical potentials
$\mu \gg T,m$ we find $\mu^* \to \mu $ and the scaling
$s\simeq c_V\propto \mu^2 T$, which implies therefore $c_V/s \to 1$.
In the degenerate regime when $T\ll \Delta$, the scaling is
$s\propto m\Delta (mT)^{1/2}e^{-\Delta/T}$.  In order to find the
leading term contributing to $c_V$, we need to keep the first thermal
correction in $\mu^*$, which gives $\mu^*\simeq m+3T/2$. Then the
integral \eqref{eq:c_v2} can be estimated as
$c_V\propto (mT)^{3/2}e^{-\Delta/T}$, and, therefore, the ratio
$c_V/s\propto T/\Delta\to 0$ as $T\to 0$ [see dashed line in panel
(c)]. In this limiting case the entropy per particle diverges as
$s/n\propto \Delta/T$, and the specific heat per particle tends to the
nonrelativistic limit $c_V/n\to 3/2$.


\providecommand{\href}[2]{#2}\begingroup\raggedright\endgroup

\end{document}